\documentclass[fleqn,usenatbib]{mnras}

\usepackage[T1]{fontenc}

\DeclareRobustCommand{\VAN}[3]{#2}
\let\VANthebibliography\thebibliography
\def\thebibliography{\DeclareRobustCommand{\VAN}[3]{##3}\VANthebibliography}

\usepackage{graphicx}	
\usepackage{amsmath}	
\usepackage{amssymb}	
\usepackage{multirow}

\usepackage{newtxtext,newtxmath}

\title[Photometric and kinematic identification of galaxy merger remnants]{The combined and respective roles of imaging and stellar kinematics in identifying galaxy merger remnants}

\author[Bottrell et al.]{
Connor Bottrell,$^{1,2}$\thanks{E-mail: connor.bottrell@ipmu.jp}
Maan H. Hani,$^{3,2}$ Hossen Teimoorinia,$^{4,2}$ \newauthor{}
David R. Patton,$^{5}$ \& Sara L. Ellison$^{2}$
\\
$^{1}$Kavli Institute for the Physics and Mathematics of the Universe (WPI), UTIAS, University of Tokyo, Kashiwa, Chiba 277-8583, Japan\\
$^{2}$Department of Physics and Astronomy, University of Victoria, Victoria, British Columbia V8P 1A1, Canada\\
$^{3}$Department of Physics and Astronomy, McMaster University, Hamilton, Ontario L8S 4M1, Canada\\
$^{4}$NRC Herzberg Astronomy and Astrophysics Research Centre, 5071 West Saanich Road, Victoria, British Columbia V9E 2E7, Canada\\
$^{5}$Department of Physics and Astronomy, Trent University, 1600 West Bank Drive, Peterborough, ON K9L 0G2, Canada\\
}

\date{Accepted XXX. Received YYY; in original form ZZZ}

\pubyear{2015}

\begin{document}
\def \nuprocess{$\nu$-process}
\def \nodata{. . .}
\def \degree{$^{\circ}$}
\def \Msolar{M$_{\odot}$}
\def \alphafe{[$\alpha$/Fe]}
\def \na{New Astronomy}
\def \HI{H\ion{i}}
\def \sion{\ion{ii}}
\def \vninety{v$_{90}$}
\def \Lbol{L$_{\rm bol}$}
\def \Mstar{M$_{\star}$}
\def \logMstar{\log($M$_{\star}/$M$_{\odot})}
\def \logRimp{log(R$_{\rm imp}$/kpc)}
\def \logLbol{log(L$_{\rm AGN}$/erg s$^{-1}$)}
\def \logsSFR{log(sSFR / yr$^{-1}$)}
\def \kms{km s$^{-1}$}
\def \zabs{z$_{\rm abs}$}
\def \zem{z$_{\rm em}$}
\def \Rimp{$\rho_{\rm imp}$}
\def \Rvir{$\rho_{\rm vir}$}
\def \deltaEW{${\rm \Delta log(EW/m\AA)}$}
\def \RadRat{$f_{\rm AGN}/f_{\rm HM01}$}
\def \mnfe{[Mn/Fe]$_{\rm DC}$}
\def \Mgeqw{W$_{0}^{2796}$}
\def \Feeqw{W$_{0}^{2600}$}
\def \fracMgFe{ ${\rm{W}_{0}^{2796}}$/${\rm{W}_{0}^{2600}}$}
\def \omegaDLA{$\Omega_{\rm H \textsc{i}}$}
\def \ndla{30}
\def \npdla{46}
\def \nlpdla{41}
\def \nxpdla{5}
\def \nmdla{27}
\def \nlmdla{21}
\def \nxmdla{6}
\def \CosmoZ{$\langle Z/Z_{\odot} \rangle$}
\def \fNX{$f(N,X)$}
\newcommand{\gimtwod}{\textsc{gim2d}}
\newcommand{\sersic}{s\'{e}rsic} 
\newcommand{\Sersic}{S\'{e}rsic}
\newcommand{\sextractor}{\textsc{SExtractor}}
\def \rifs{\textsc{RealSim-IFS}}
\def \realsim{\textsc{RealSim}}
\def \tpost{T_{\mathrm{PM}}}

\label{firstpage}
\pagerange{\pageref{firstpage}--\pageref{lastpage}}
\maketitle

\begin{abstract}
One of the central challenges to establishing the role of mergers in galaxy evolution is the selection of pure and complete merger samples in observations. In particular, while large and reasonably pure interacting galaxy pair samples can be obtained with relative ease via spectroscopic criteria, automated selection of post-coalescence merger \emph{remnants} is restricted to the physical characteristics of remnants alone. Furthermore, such selection has predominantly focused on imaging data -- whereas kinematic data may offer a complimentary basis for identifying merger remnants. Therefore, we examine the theoretical utility of \emph{both} the morphological and kinematic features of merger remnants in distinguishing galaxy merger remnants from other galaxies. Deep classification models are calibrated and evaluated using \emph{idealized} synthetic images and line-of-sight stellar velocity maps of a heterogeneous population of galaxies and merger remnants from the TNG100 cosmological hydrodynamical simulation. We show that \emph{even} idealized stellar kinematic data has limited utility compared to imaging and under-performs by $2.1\%\pm0.5\%$ in completeness and $4.7\%\pm0.4\%$ in purity for our fiducial model architecture. Combining imaging and stellar kinematics offers a small boost in completeness (by $1.8\%\pm0.4\%$, compared to $92.7\%\pm0.2\%$ from imaging alone) but no change in purity ($0.1\%\pm0.3\%$ improvement compared to $92.7\%\pm0.2\%$, evaluated with equal numbers of merger remnant and non-remnant control galaxies). Classification accuracy of all models is particularly sensitive to physical companions at separations $\lesssim40$ kpc and to time-since-coalescence. Taken together, our results show that the stellar kinematic data has little to offer in compliment to imaging for merger remnant identification in a heterogeneous galaxy population. 
\end{abstract}




\section{Introduction}

Galaxy mergers have a fundamental and critical role within the Lambda cold dark matter ($\Lambda$CDM) concordance cosmogony \citep{1978MNRAS.183..341W,1991ApJ...379...52W}. In this paradigm, large structures form through continuous and diverse merging events between smaller structures (e.g. \citealt{1993MNRAS.262..627L}). However, the role of mergers in galaxy formation and evolution is not limited to the \emph{ex-situ} assembly of mass. Theoretical models and numerical simulations show that the strong gravitational and tidal forces in mergers can be responsible for morphological and dynamical transformation of galaxies (e.g. \citealt{1972ApJ...178..623T,1977egsp.conf..401T,1983MNRAS.205.1009N,1985Natur.317..595F,1992ApJ...400..460H,2002NewA....7..155S,2003ApJ...597..893N,2005A&A...437...69B,2007MNRAS.376..997J,2009ApJ...691.1168H,2014MNRAS.444.3357N}). These theoretical results are supported by the disrupted morphologies (e.g. \citealt{2005AJ....130.2043P,2005AJ....129..682H,2006AJ....132...71H,2007AJ....134.2286H,2007ApJ...666..212D,2010MNRAS.407.1514E,2013MNRAS.429.1051C,2016MNRAS.461.2589P,2021arXiv210501675W}) and spatially-resolved kinematic disturbances (e.g. \citealt{2015A&A...582A..21B,2015ApJ...803...62H,2017MNRAS.465..123B,2018MNRAS.476.2339B,2020A&A...634A..26P,2020ApJ...892L..20F,2021MNRAS.504.1989W}) that are often observed in interacting galaxies and recent merger remnants (\emph{post-mergers}) relative to non-merging galaxies. Transformation of galaxy morphologies and kinematics through interactions and mergers is qualitatively consistent with the observational framework relating morphology and kinematics to galaxy environment (e.g. \citealt{1974ApJ...194....1O,1976ApJ...208...13D,1980ApJ...236..351D,2009MNRAS.393.1324B,2011MNRAS.416.1680C,2019MNRAS.485..666B}). 

Interactions are also connected to a number of internal processes. Compared to non-interacting galaxies, observed merging galaxy pairs and post-mergers exhibit enhanced central star formation rates (e.g. \citealt{Barton_2000,2007ApJ...660L..51L,2008AJ....135.1877E,2013MNRAS.435.3627E,2012MNRAS.426..549S,2013MNRAS.433L..59P,2019MNRAS.482L..55T,2019ApJ...881..119P,2020MNRAS.493.3716H,2021MNRAS.503.3113M}), redistributed gas content (e.g. \citealt{2006AJ....131.2004K,2008AJ....135.1877E,2013MNRAS.435.3627E,2008MNRAS.386L..82M,2010ApJ...723.1255R,2019MNRAS.482L..55T,2019MNRAS.485.1320M}), and increased incidence of AGN (e.g. \citealt{2007MNRAS.375.1017A,2011MNRAS.418.2043E,2019MNRAS.487.2491E,2011ApJ...743....2S,2014MNRAS.441.1297S,2018PASJ...70S..37G}). 

However, the \emph{relative} role of galaxy mergers in these morphological, dynamical, and internal processes and their sensitivities to merger stage and initial conditions are currently uncertain. For example, what is the relative contribution of mergers to triggering star formation and AGN compared to the contribution from accretion of cold gas along cosmic filaments (e.g. \citealt{2006MNRAS.368....2D,2007MNRAS.380..339B,2009Natur.457..451D,2011ApJ...741L..33B,2020MNRAS.494.5713M})? What is their relative role in the structural and dynamical evolution of galaxies? Is the relative role of mergers in these processes static or does it evolve with cosmic time? Ultimately, robustly characterizing the role of mergers in galaxy evolution requires (1) statistically representative or bias-corrected censuses of observed galaxy mergers and (2) the ability to link galaxy mergers to their stages and initial conditions (stellar mass ratio and orbital parameters of the merging galaxies, their gas fractions, and morphologies). Both of these criteria present significant challenges from an observational perspective -- particularly for the post-merger phase when spectroscopic validation is not possible. 

\begin{figure*}
\centering
	\includegraphics[width=0.493\linewidth]{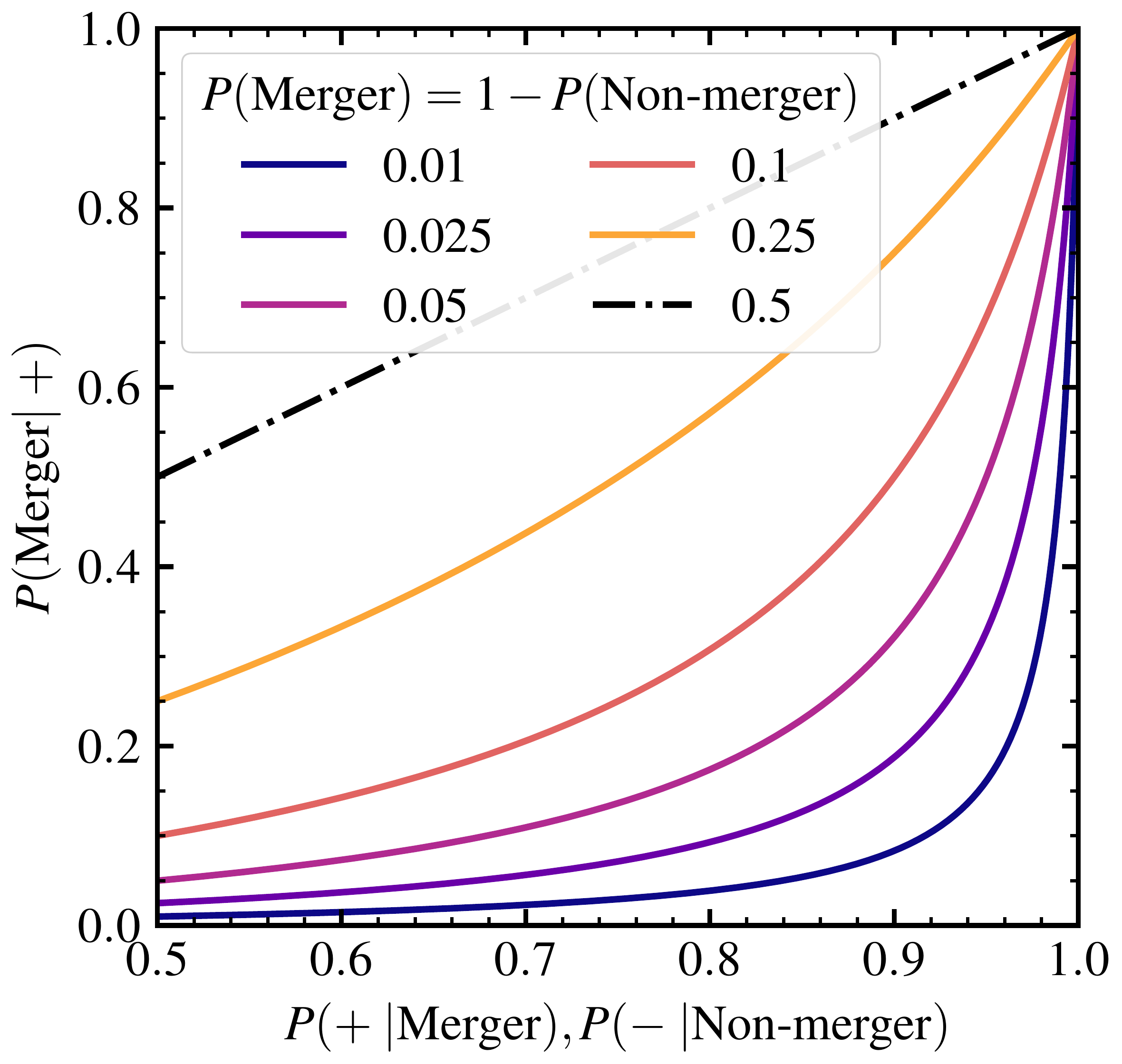}
	\includegraphics[width=0.499\linewidth]{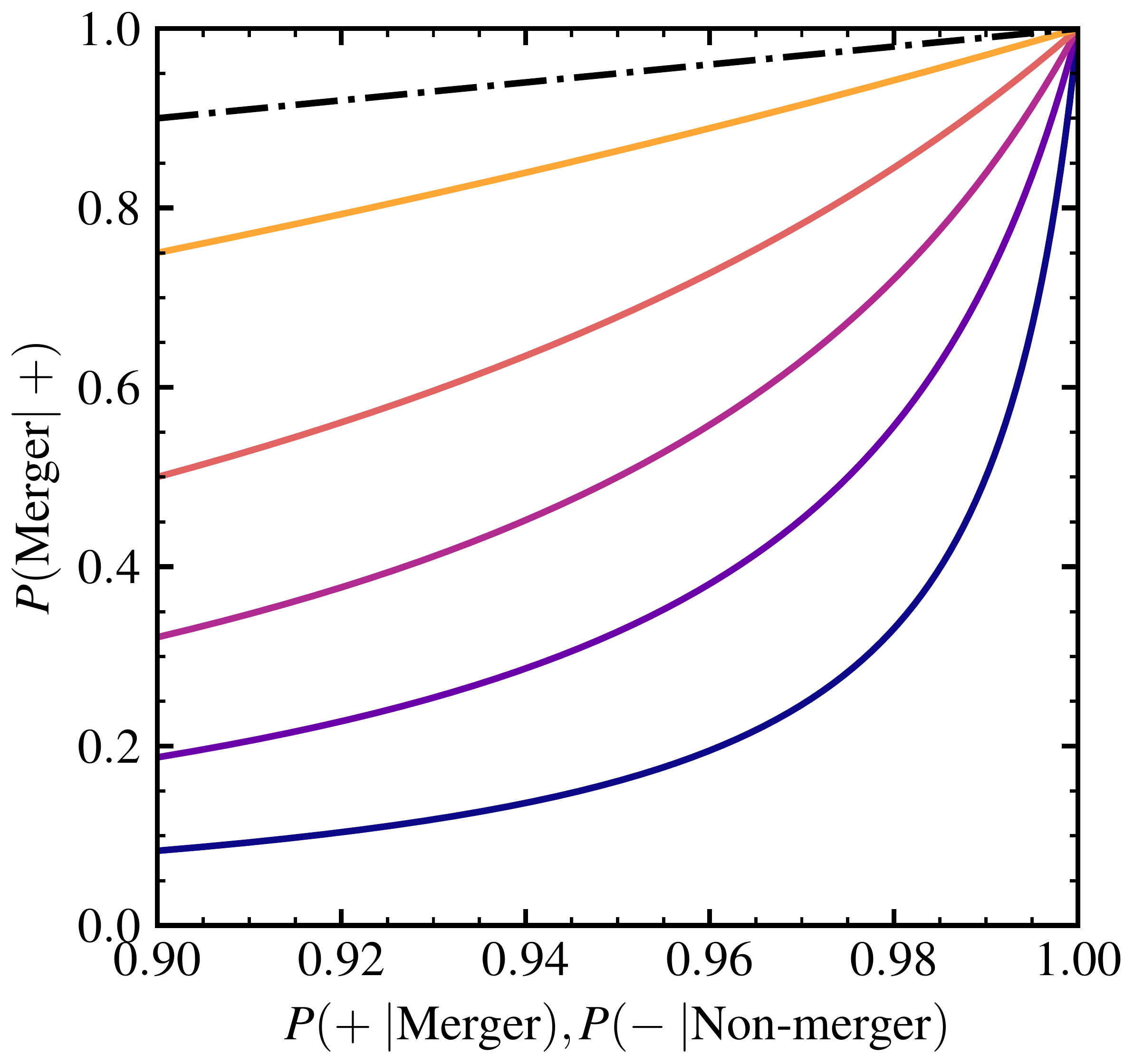}
    \caption[Bayes' theorem in a hypothetical classifier]{An illustration of the role of priors in Bayesian statistical inference with a hypothetical binary merger/non-merger classification model. The left panel shows the relationship between the posterior probability, $P(\mathrm{Merger}|+)$, and a model's sensitivity (completeness), $P(+|\mathrm{Merger})> 0.5$, for a range of of priors, $P(\mathrm{Merger})$, plotted as dash-dot and coloured curves. For simplicity, the specificity of this hypothetical model is equal to its sensitivity, $P(-|$Non-merger$)=P(+|\mathrm{Merger})$, and the two groups are complementary, $P($Merger$) = 1-P($Non-merger$)$. The right panel zooms-in on sensitivities greater than 90\% for better contrast in this regime.}
    \label{fig:bayes}
\end{figure*}

Models that can quantitatively classify and characterize galaxy mergers in an automated way present an attractive solution to these challenges because they overcome the potential for ambiguity in visual classifications and can rapidly process large samples of galaxies (e.g. \citealt{1994ApJ...432...75A,2003ApJS..147....1C,2004AJ....128..163L,2014ApJ...787..130W,2015ApJ...803...62H,2016MNRAS.456.3032P}; McElroy et al. in prep). In particular, automated merger classification models can be calibrated based on synthetic observations of galaxies and galaxy mergers from hydrodynamical simulations (e.g. \citealt{2008MNRAS.391.1137L,2010MNRAS.404..590L,2010MNRAS.404..575L,2019A&A...626A..49P,2019ApJ...872...76N,2019MNRAS.490.5390B,2020A&C....3200390C,2021arXiv210301373C,2021MNRAS.506..677C,2021MNRAS.504..372B,2020ApJ...895..115F,2021ApJ...912...45N}; McElroy et al. in prep). The advantage of this approach is that the simulations provide \emph{a priori} knowledge of whether galaxies are undergoing interactions as well as the stage and initial conditions of the interactions. Therefore, the biases and limitations in estimating these parameters for observed galaxies do not become embedded in the calibration of the models. Indeed, the features which observers use to visually distinguish merging galaxies from non-merging galaxies (such as stellar tails, tidal arms, bridges, and shells, and kinematic asymmetries) are \emph{motivated} by the results of numerical experiments (e.g. \citealt{1972ApJ...178..623T,1989ApJ...342....1H,1992ARA&A..30..705B,1998ApJ...494..183M,1999MNRAS.307..495H,2005ApJ...621..725C,2008ApJ...689..936J,2020MNRAS.492.2075B}). Consequently, an automated merger classification model that is calibrated on realistic synthetic observations from simulations will be optimized to distinguish mergers from non-mergers based on the same set of features that are expected to appear in observations -- insofar as the simulations are realistic. 

\subsection{The importance of purity in merger remnant identification}
Nevertheless, an automated merger classification model that can generate pure and complete merger samples remains elusive. Despite their variety and advantages, automated classifiers must observe statistical laws. In particular, the statistical prior from the low incidence of observed on-going mergers and remnants, $P(\mathrm{Merger})$ (i.e. true merger fraction), presents one of the greatest challenges to automated merger classification. Figure \ref{fig:bayes} illustrates the role of statistical priors in a hypothetical binary classification model. $P(+|$Merger$)$ is the probability of a positive classification for a merger (\emph{completeness}, true positive rate, or \emph{sensitivity}). $P(-|$Non-merger$)$ is the probability of a negative result for a non-mergers (true negative rate, or \emph{specificity}). The sensitivity and specificity are assumed to be equal for simplicity. The corresponding false positive rate (fall-out) and false negative rate (miss rate) of the model are then $P(+|$Non-merger$) = 1-P(-|$Non-merger$)$ and $P(-|$Merger$) = 1-P(+|$Merger$)$. From \cite{bayes1763lii}, the posterior probability that a galaxy classified as $+$ is actually a merger is then:


\begin{align}
\begin{split}\label{eq:bayes}
{}&P(\mathrm{Merger}|+) = \frac{P(+|\mathrm{Merger})P(\mathrm{Merger})}{P(+)}\\ &  = \frac{P(+|\mathrm{Merger})\;P(\mathrm{Merger})}{P(+|\mathrm{Merger})\;P(\mathrm{Merger}) + P(+|\text{Non-merger})\;P(\text{Non-merger})}
\end{split}
\end{align}
which can also be interpreted as the \emph{expectation value for the purity} of the out-falling merger sample produced by the model. Figure \ref{fig:bayes} shows $P($Merger$|+)$ as a function of sensitivity for a range of prior incidence probabilities, $P($Merger$)$ -- plotted as dot-dashed and coloured curves. The right panel zooms-in on $P(+|$Merger$)>0.9$. Figure \ref{fig:bayes} shows that a decrease in the underlying incidence of mergers strongly effects the posterior probabilities and corresponding sample purities. For example, even with $99\%$ sensitivity and specificity but prior incidence of 1\% (navy curve), the corresponding posterior probability that a galaxy is actually a merger of Group A given that it has a positive ($+$) result from the model is only $50\%$.

The selection functions for galaxy mergers vary greatly -- both in terms of merger definition and selection methods. For reference, typical merger fraction estimates in the nearby Universe ($z\lesssim0.5$) are in the range of $\sim1-10\%$ (e.g. \citealt{2004ApJ...617L...9L,2007ApJS..172..320K,2008ApJ...681..232L,2008ApJ...685..235P,2009A&A...498..379D,2011ApJ...742..103L,2012A&A...548A...7L}). For a merger classification model that is 99\% sensitive and specific, these priors on merger fractions would yield corresponding posterior probabilities of $50-92\%$ (expected purities). Figure \ref{fig:bayes} also shows that these purities fall rapidly with decreasing sensitivity. At $P(+|$Merger$) = P(-|$Non-merger$)$ = 0.9, galaxies that are classified as positive ($+$) by the model only have $8-50\%$ probabilities of actually being a merger. In other words, the sample of ``mergers'' classified by the model would realistically comprise $8-50\%$ true mergers and, respectively, $92-50\%$ contaminating non-merger galaxies. In particular, \cite{2021MNRAS.504..372B} explicitly show that such low purities in merger samples (high contamination rates) result in significant under-estimates for the impact of mergers in driving physical changes in galaxies (e.g. enhanced star-formation rates; Figure 16 in \citealt{2021MNRAS.504..372B}).

\subsection{Rationale for multi-variate classification with kinematics}
Presently, automated merger classification and characterization models are predominantly based on optical morphologies or other information derived from imaging (e.g. recently \citealt{2016MNRAS.456.3032P,2018MNRAS.479..415A,2019MNRAS.486.3702S,2019MNRAS.483.2968W,2019A&A...626A..49P,2019ApJ...872...76N,2019MNRAS.490.5390B,2020A&C....3200390C,2021MNRAS.506..677C}). However, classification performance may be improved using other information from which a model can draw -- independently, or in tandem with imaging. In particular, mergers and recent post-mergers \emph{can} exhibit complex kinematics compared to non-merging galaxies (e.g. \citealt{1998ApJ...494..183M,2005ApJ...621..725C,2008ApJ...682..231S,2015ApJ...803...62H,2018MNRAS.476.2339B,2020ApJ...892L..20F}) which are sensitive to interaction stage and other merger characteristics (\citealt{2016ApJ...816...99H,2021ApJ...912...45N}). Importantly, kinematics and morphology can tell different stories about the state of a galaxy. Merging galaxies and post-mergers can appear morphologically ordinary for certain orientations or in images that are too shallow to reveal their low surface-brightness tidal signatures \citep{2008MNRAS.391.1137L,2010MNRAS.404..575L,2014A&A...566A..97J,2019MNRAS.486..390B}. Such galaxies (particularly post-mergers) may still show elevated disturbances in their kinematics \citep{2016ApJ...816...99H}. Similarly, non-merging galaxies with high star formation rates can be morphologically asymmetric or ``clumpy'' but kinematically regular (e.g. \citealt{2013PASA...30...56G,2019ApJ...874...59S}). Galaxy stellar kinematics may therefore be complementary to imaging from the perspective of merger characterization and may help to overcome the challenges of building modules which produce merger samples with high completeness and purity.

Highly-multiplexed integral field spectroscopy (IFS) surveys have generated a revolutionary expansion in the volume and quality of spatially-resolved galaxy kinematic data in the local Universe (e.g. the Calar Alto Legacy Integral Field Area (CALIFA): 600 galaxies \citealt{2012A&A...538A...8S}, the Sydney-AAO Multi-object IFS survey (SAMI): 3,600 galaxies \citealt{2012MNRAS.421..872C}, the Mapping Nearby Galaxies at Apache Point Observatory survey (MaNGA): 10,000 galaxies \citealt{2015ApJ...798....7B}). Indeed, the growth and quality of the data are expected to continue with forthcoming instruments (e.g. Hector: 15,000 galaxies \citealt{2018SPIE10702E..1HB}). Consequently, galaxy kinematics have finally become both attractive and \emph{viable} for building pure and reasonably complete merger samples that can subsequently be further characterized. 

In this paper, we investigate the degree to which galaxy kinematics stand to improve the purity and completeness with which mergers can be detected in large IFS surveys. In particular, we focus on galaxy merger remnants where, unlike the interacting pair phase, spectroscopic validation from the redshifts of companions is not possible. To do so, we calibrate and evaluate deep convolutional classification models on idealized, high-resolution, high-dispersion, synthetic stellar velocity maps and corresponding stellar mass surface density images of galaxies from the TNG100-1 cosmological hydrodynamical simulation \citep{2018MNRAS.475..624N,2018MNRAS.475..648P}. Our approach of using idealized synthetic data is to first evaluate whether more sensitive and specific post-merger identification could be achieved by incorporating stellar kinematic data \emph{even in the absence of instrumental limitations} for diverse and statistically generalizable galaxy/merger populations such as are found in TNG100-1. Then, if the kinematic information proves integral to post-merger classification, synthetic kinematic data can be justifiably produced with more detailed and realistic accounting of observational and instrumental effects (e.g. \citealt{2019MNRAS.490.5390B,2021ApJ...912...45N}).

This paper is laid out as follows. Section \ref{sec:methods} describes: (\ref{sec:tng}) the salient features of the TNG100-1 simulation; (\ref{sec:moments}) construction of idealized synthetic images and stellar kinematic maps, (\ref{sec:selection}) the selection of post-merger galaxies and a corresponding non-post-merger reference sample; and (\ref{sec:models}) the architectures and calibration of our classification models. Our results are presented in Section \ref{sec:results} and discussed in Section \ref{sec:discussion}. Our conclusions are summarized in Section \ref{sec:conclusions}. The calculations in this work adopt Planck-based cosmological parameters with $H_0 = 67.7$ km s$^{-1}$ Mpc$^{-1}$, $\Omega_{\mathrm{m}} = 0.307$, and $\Omega_{\Lambda}=0.693$ \citep{2016A&A...594A...1P}.

\section{Methods}\label{sec:methods}

In this section, we (1) summarize the salient features of the TNG100-1 simulation and describe the synthetic kinematic and imaging data, (2) describe our selection of post-merger galaxies and corresponding non-post-merger controls, and (3) describe our post-merger galaxy classification models and the procedures used to calibrate and evaluate them. 
 
\subsection{Simulations and synthetic data}

\subsubsection{TNG100-1 Simulation}\label{sec:tng}

We use the publicly available data from the IllustrisTNG simulations\footnote{\href{https://www.tng-project.org/data/}{https://www.tng-project.org/data}} to calibrate and evaluate post-merger galaxy classification models using stellar kinematics and imaging. IllustrisTNG is suite of large-volume cosmological magneto-hydrodynamical simulations of galaxy formation and evolution \citep{2018MNRAS.479.4056W,2018MNRAS.475..648P,2018MNRAS.473.4077P,2018MNRAS.475..676S,2018MNRAS.477.1206N,2018MNRAS.480.5113M,2018MNRAS.475..624N,2019MNRAS.490.3234N,2019ComAC...6....2N}. The simulations track the evolution of dark matter, gas, stars, supermassive black holes, and magnetic fields from redshift $z=127$ to $z=0$ using the AREPO moving-mesh hydrodynamic code \citep{2010MNRAS.401..791S}. The simulations are run in three cubic volumes $(51.7^3,\;110.7^3,\;302.6^3)$ cMpc$^3$ with descending levels of mass and effective spatial resolution. All of the runs use the same physical model described by \cite{2018MNRAS.479.4056W} and \cite{2018MNRAS.475..648P} -- which builds on the original Illustris model \citep{2013MNRAS.436.3031V,2014MNRAS.438.1985T,2014Natur.509..177V,2014MNRAS.445..175G}. We adopt the fiducial run for the $110.7^3$ cMpc$^3$ volume (TNG100-1) for our investigation. The TNG100-1 volume comprises a statistically representative sample of galaxies with realistic and diverse morphologies and kinematics (e.g. \citealt{2019MNRAS.483.4140R, 2019MNRAS.487.5416T, 2019MNRAS.489.1859H, 2020arXiv200204182D}). These features of TNG100-1 make it highly appropriate for calibrating merger-detection models. The simulation's effective baryonic spatial resolution is approximately $1$ kpc. The merger trees of TNG100-1 galaxies are derived using the \textsc{SubLink} algorithm \citep{2015MNRAS.449...49R} -- which links a given galaxy  to its progenitors and descendants. We also use a wealth of ancillary information derived from the IllustrisTNG group catalogues by \cite{2020arXiv200300289P} and \cite{2020MNRAS.493.3716H}, including environmental information and merger properties. 

\subsubsection{Synthetic LOSVD Moment Maps} \label{sec:moments}

Synthetic stellar line-of-sight-velocity distribution (LOSVD) cubes were produced for all galaxies in TNG100-1 with redshifts $z\leq1$ (snapshots $\geq 50$) and subhalo stellar masses $\logMstar{}\geq10$ for a total of $303,110$ galaxies\footnote{The $\logMstar{}\geq10$ stellar mass cut is chosen such that each galaxy in our sample comprises at least $7000$ stellar particles and, consequently, their stellar structures are at least reasonably well-resolved.} The cubes were produced along four lines of sight for each galaxy with an adaptive field-of-view set to 10 times the galaxy stellar half-mass radius with 4.67 km/s velocity resolution and $\pm700$ km/s velocity grid limits. The positions and velocities of stellar particles within 20 stellar half-mass radii relative to the galaxy's gravitational potential minimum are deposited onto the $(x,y,v)_{\mathrm{los}}$ grid (i.e. systemic velocities removed). Stellar particles are smoothed in $(x,y)_{\mathrm{los}}$ using an adaptive cubic spline kernel \citep{1992ARA&A..30..543M} with smoothing length for each stellar particle set to its 3-dimensional $32^{\mathrm{nd}}$ nearest-neighbour distance. Each stellar particle is assumed to have a single discrete line-of-sight velocity and we do not account for internal velocity dispersion of stellar particles. The final cubes have dimensions $(N_x,N_y,N_v)_{\mathrm{los}}=(512,512,300)$.

Line-of-sight stellar mass surface density, mass-weighted velocity, and velocity dispersion maps (stellar LOSVD moment maps) are derived independently from the LOSVDs in every spaxel of the cubes without binning. Therefore, due to the adaptive smoothing we employ, even spaxels with arbitrarily low stellar surface densities have measurements. Unbiased estimates of the stellar LOSVD moments are computed in each spaxel as follows\footnote{See \citet{2014A&C.....5....1R} for comprehensive derivations of the \emph{unbiased} estimators for the moments of weighted samples.}:
\begin{align}
\Sigma_{\star} =&\frac{1}{p^2} \sum_{k=1}^{N_v} m_k \quad \text{where $p$ is the pixel scale (kpc);} \label{eq:moment0}\\ 
\bar{v} =&\frac{1}{V_1}\sum_{k=1}^{N_v}m_k v_k \quad \text{where $V_1=\sum_{k=1}^{N_v} m_k; \quad$ and}\label{eq:moment1}\\
\sigma^2 =& \frac{V_1}{V_1^2-V_2} \sum_{k=1}^{N_v} m_k (v_k - \bar{v})^2 \quad \text{where $V_2=\sum_{k=1}^{N_v} m^2_k$\label{eq:moment2}}
\end{align}
where $m_k$ is the mass weight in velocity element $v_k$, and where we have implicitly assumed a single component for the LOSVD in each spaxel. Our analysis focuses on (Eq. \ref{eq:moment0}) the stellar mass surface density maps for their strong resemblance to photometry and (Eq. \ref{eq:moment1}) the line-of-sight velocity maps for the stellar kinematics.  

Each individual stellar mass surface density map is logarithmically scaled to optimize the contrast of high and low-surface brightness features and result in images with $[0,1]$ intensities. Similarly, the velocity maps are scaled linearly to [-1,1] using $\min(|v_{0.1\%}|, |v_{99.9\%}|)$ and clipping outliers.

\begin{figure}
\hspace{-4pt}
	\includegraphics[width=\linewidth]{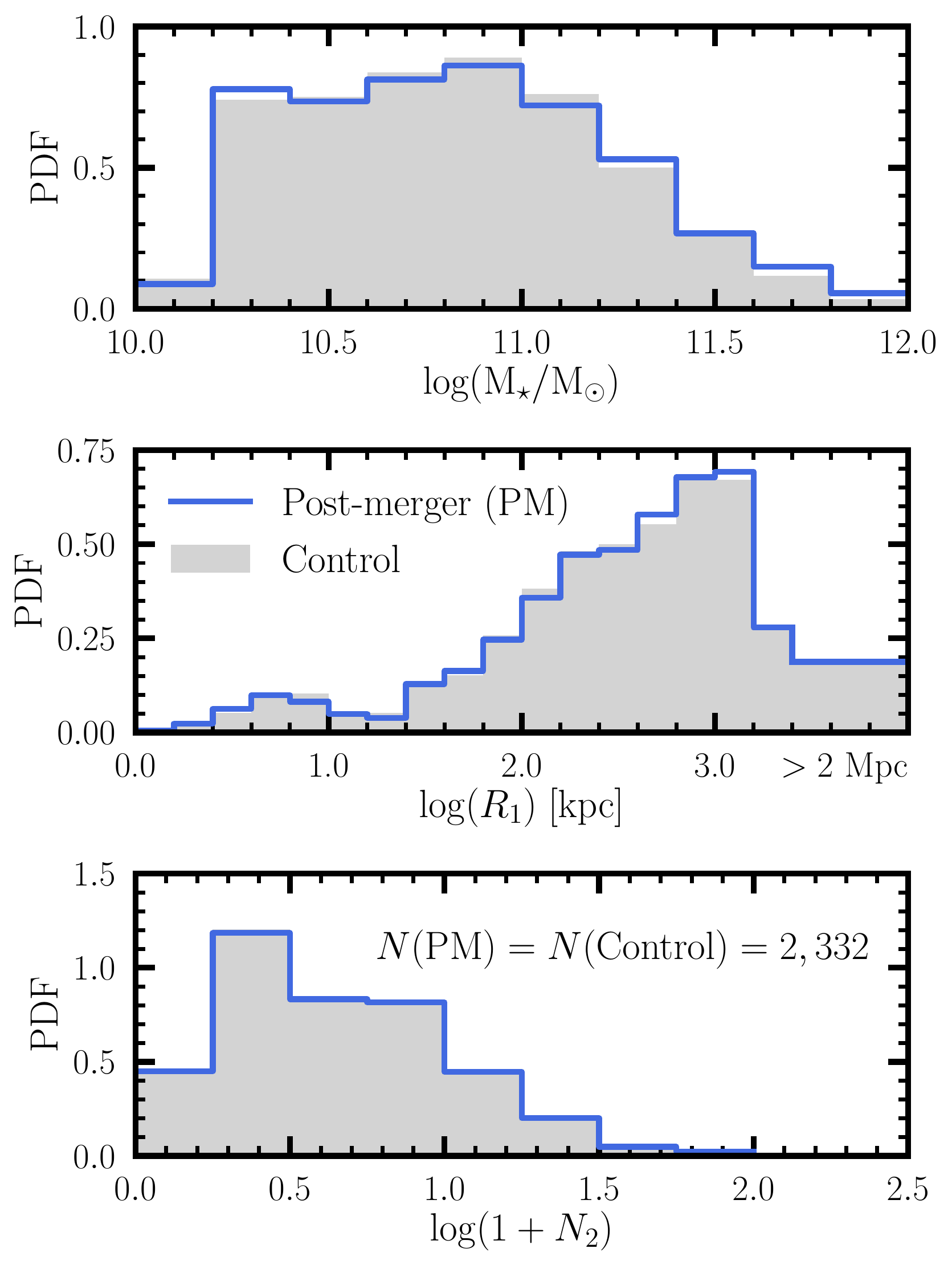}
   \caption[Post-merger and control selection]{Post-merger and matching non-post-merger control galaxy samples. Each post-merger is matched to a non-post-merger control galaxy ($\tpost>2$ Gyr$)$ on redshift, stellar mass, and environment. Environmental parameters $R_1$ and $N_2$ only consider neighbours whose stellar mass is at least a tenth of the stellar mass of the target galaxy. The large hatched bin in the $R_1$ panel (middle) is for galaxies for which no $M_{\star,2} \geq 0.1 M_{\star,1}$ neighbour was found within 2 Mpc.}
    \label{fig:selection}
\end{figure}

\begin{figure*}
\hspace{-4pt}
	\includegraphics[width=0.495\linewidth]{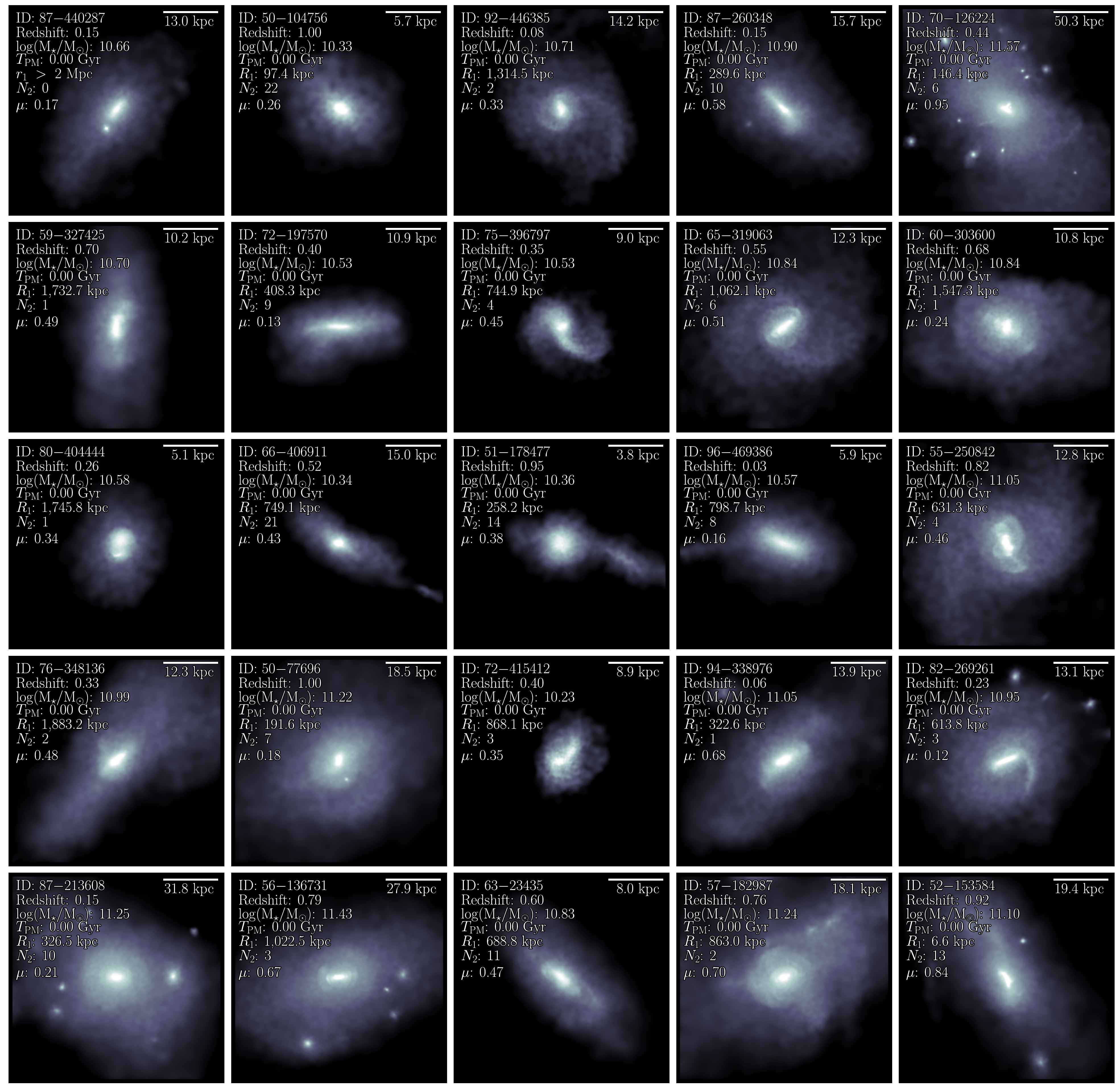}
	\includegraphics[width=0.495\linewidth]{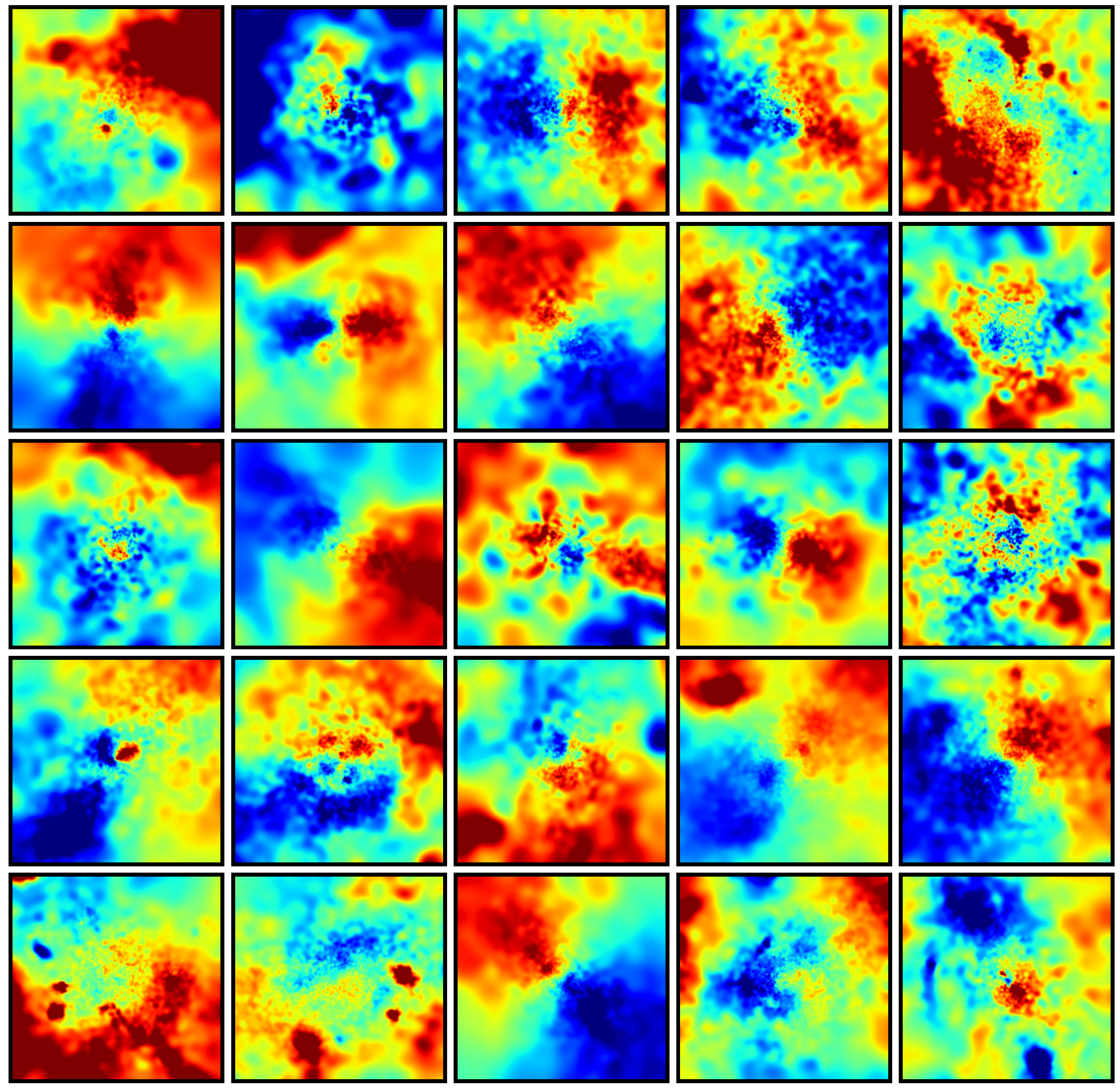}\vspace{1pt}
	\includegraphics[width=0.495\linewidth]{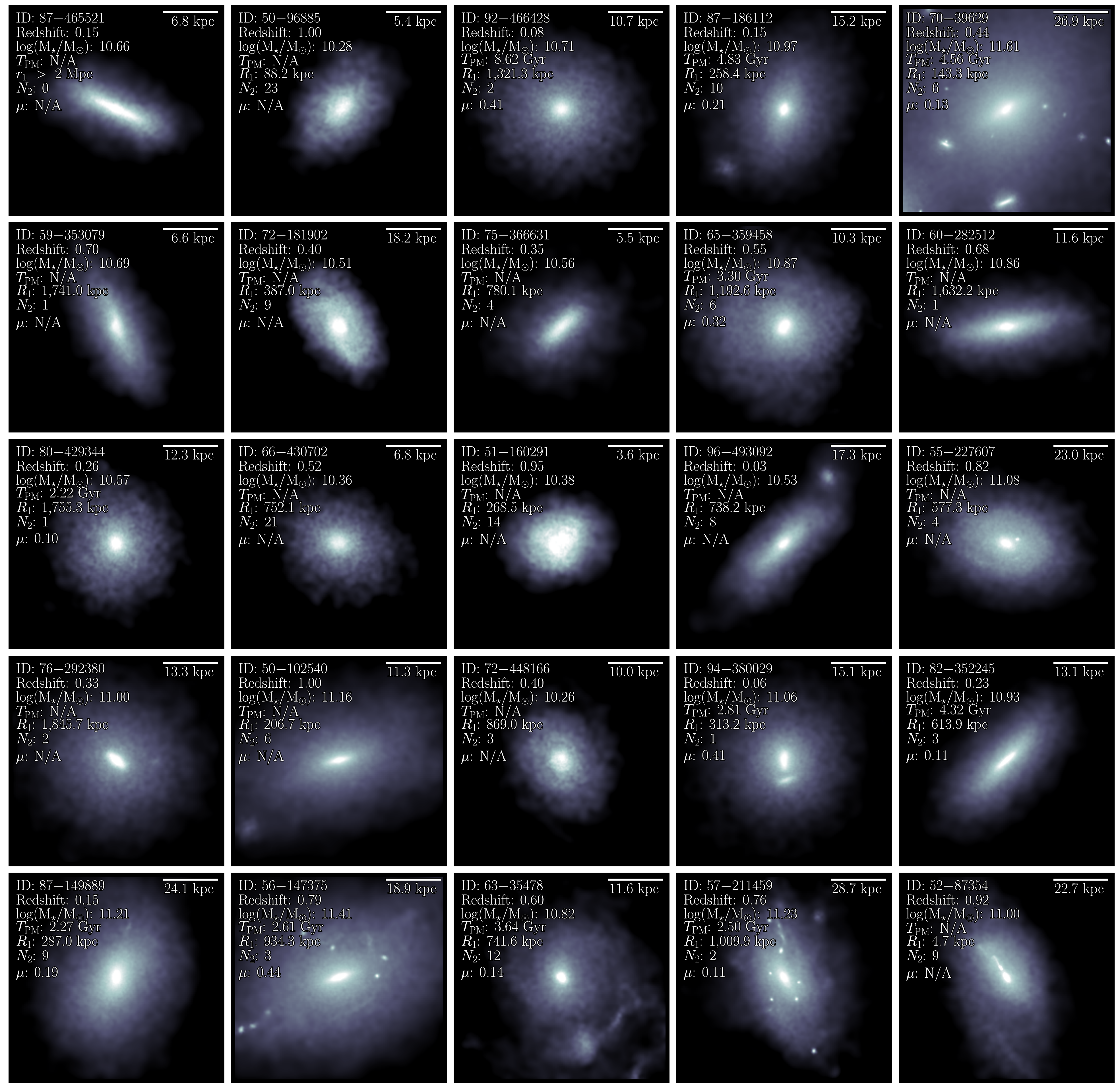}
	\includegraphics[width=0.495\linewidth]{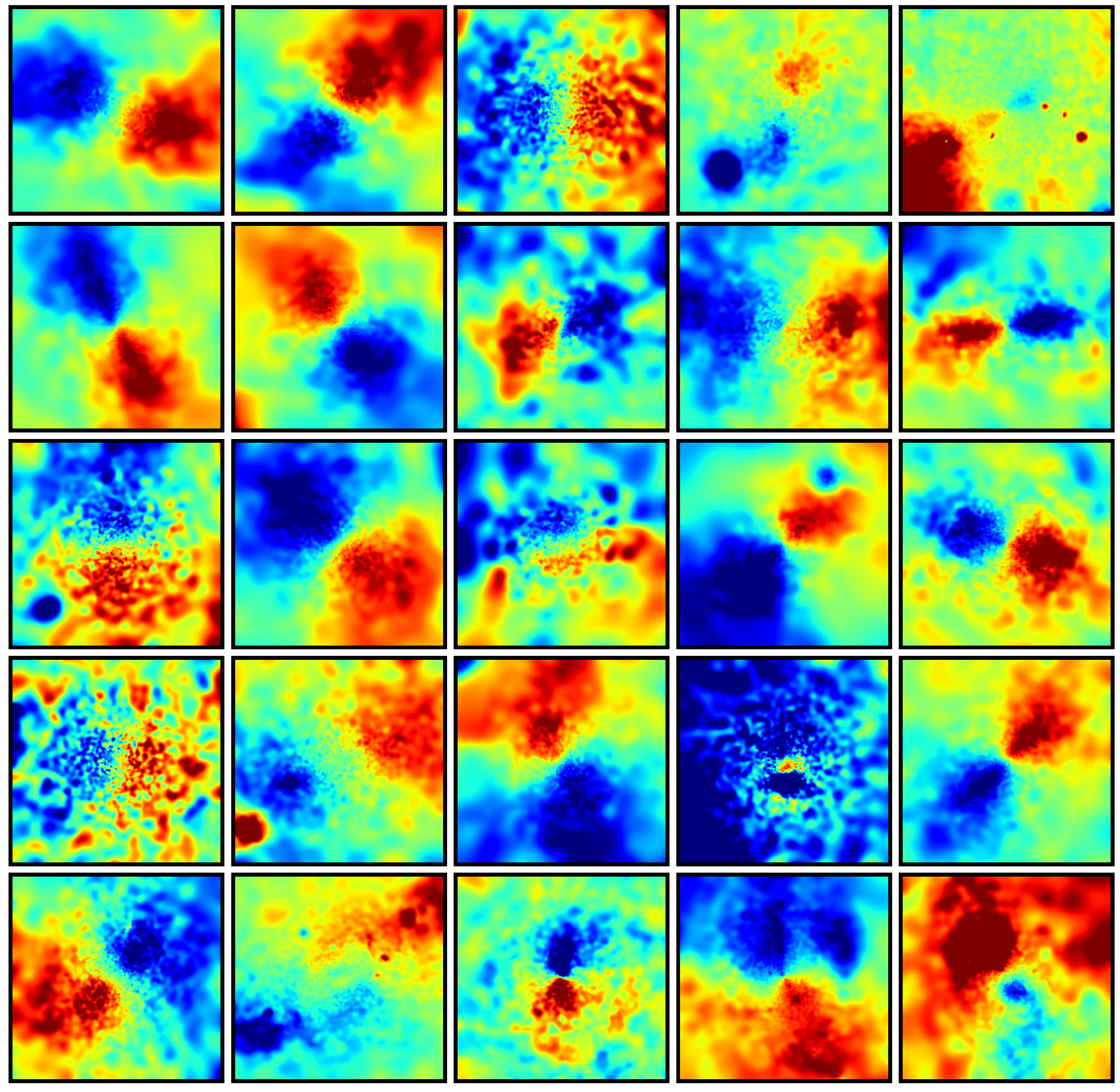}
   \caption[Idealized LOSVD moment maps]{Stellar surface density maps (left panels) and matching line-of-sight velocity maps (right panels) for TNG100-1 galaxies randomly selected from our post-merger sample (upper panels) and their corresponding matched non-post-merger control galaxies (lower panels). Only one of four lines of sight for each galaxy is shown. \emph{On-line version only}: Galaxy IDs labeled in each panel comprise their $\{\texttt{SnapNum}\}$-$\{\texttt{SubhaloID}\}$. Non-post-merger galaxies with $\mu=0$ have had no merger with $\mu\geq0.1$ since our merger tree redshift limit of $z=2$ and their $\tpost$ values are the corresponding time since $z=2$.}
    \label{fig:mosaics}
\end{figure*}

\subsection{Post-merger and Control Selection}\label{sec:selection}
For every galaxy in our sample, \cite{2020MNRAS.493.3716H} used the \textsc{SubLink} merger trees to compute the time since the most recent merger (post-merger time), $\tpost$, in which the stellar mass ratio of the first (most massive) and second progenitors were $\mu = M_{\star,2}/M_{\star,1} \geq0.1$ where $M_{\star,1} \geq M_{\star,2}$ by definition. The maximum mass enclosed within 2 stellar half-mass radii of each progenitor galaxy in the last 500 Myr is used to compute $\mu$, which is done to avoid numerical stripping effects in the estimate of $\mu$ when progenitors have overlapping stellar components (e.g. see \citealt{2015MNRAS.449...49R,2020arXiv200300289P}). For our post-merger sample, we take all galaxies from our main sample described in \ref{sec:tng} with $\tpost=0$ (i.e. galaxies that are immediate post-merger remnants of two or more progenitors in the previous simulation snapshot). This selection yields a post-merger sample of $2,332$ galaxies -- each of which is observed along four lines of sight as described in Section \ref{sec:moments}. This post-merger sample is identical to the one used for post-merger classifications by \cite{2021MNRAS.504..372B}.

As in \cite{2021MNRAS.504..372B}, a reference sample of non-post-merger control galaxies is constructed which match the post-merger galaxies in redshift, stellar mass, and environment. This rigorous control matching is intended to eliminate the influence of exploitable correlations between these properties and merger status and also result in calibrated classification probabilities (\citealt{10.1145/1102351.1102430}). The control pool comprises galaxies from the main sample whose $\tpost>2$ Gyr. We then create a balanced post-merger/control dataset by selecting the best-matching control for each post-merger. For a given post-merger, the first control criterion is that it belongs to the same redshift snapshot. Then, the best-matching control in stellar mass, $M_{\star}$, nearest neighbour distance, $R_1$, and number of galaxies within 2 Mpc, $N_2$, is taken from the control pool without replacement\footnote{In practice, an iterative process was used to identify the best-matching \emph{set of controls} for each galaxy within a strict $0.1$ dex ($10\%$) tolerance modified by an integer growth factor in each iteration. From this matched set, the best-matching control was extracted. The mean growth factor was $N_{\mathrm{grow}} = 1.77$. The majority of controls, $58\%$, are matched with the unmodified tolerance at $N_{\mathrm{grow}} = 1$ and 84\% are matched by $N_{\mathrm{grow}} = 2$. This control sample is \emph{not strictly identical} to the one used in \cite{2021MNRAS.504..372B} as different matching algorithms were used.}. For the environmental parameters $R_1$ and $N_2$, only galaxies whose stellar masses are at least $0.1 M_{\star}$ are considered. Figure \ref{fig:selection} shows the distributions of $M_{\star}$, $R_1$, and $N_2$ for the post-merger and matched control samples.

Figure \ref{fig:mosaics} shows the stellar mass surface density and velocity maps for 25 randomly selected post-merger galaxies (upper panels) and their corresponding matched controls (lower panels). Control-matching parameters, ($z$, $M_{\star}$, $R_1$, and $N_2$)  and other post-merger properties ($\mu$ and $\tpost$) are shown in the upper left of each map (online version only). Figure \ref{fig:mosaics} shows the diversity of the post-merger and control samples. Most post-mergers in Figure \ref{fig:mosaics} frequently exhibit asymmetric structures in their morphologies typical of tidal interactions -- but not always. Meanwhile, the control sample \emph{predominantly} exhibits more ``normal" morphologies, but also includes irregular and asymmetric morphologies which may be driven by environment, secular evolution, or more minor interactions with $\mu<0.1$  (e.g. \citealt{1994A&A...290L...9R,1995ApJ...447...82R,10.1046/j.1365-8711.1999.02332.x,2009PhR...471...75J,2013ApJ...772..135Z,2018MNRAS.480.2266M,2021arXiv210908882Y}). Ongoing $\mu\geq0.1$ interactions and fly-bys are also possible within the control sample via our environmental matching criteria. Calibration of a proper post-merger classification model using simulations must account for this overlapping of the post-merger and non-post-merger feature-spaces. 

Similarly, Figure \ref{fig:mosaics} shows that while several controls exhibit complex stellar kinematics, high kinematic asymmetries are more common amongst the post-mergers shown. In particular, a few post-mergers with seemingly regular morphologies have disturbed kinematics which betray their post-merger status (e.g. upper panels [row,column]: [1,2] and [1,4]). Meanwhile, some of the more morphologically irregular controls exhibit regular kinematics (e.g. lower panel: [2,2] and [5,3]). It is unclear from these kinematic maps whether they alone can be used to calibrate a classification model that will yield complete and pure post-merger samples. However, they may be complimentary when used in tandem with imaging. Therefore, several model architectures for post-merger classification are considered: those which use the stellar surface density and kinematic maps separately, and multi-input models which incorporate both datasets together.

\subsection{Post-merger Classification Models}\label{sec:models}

\begin{figure}
\hspace{-4pt}
	\includegraphics[width=\linewidth]{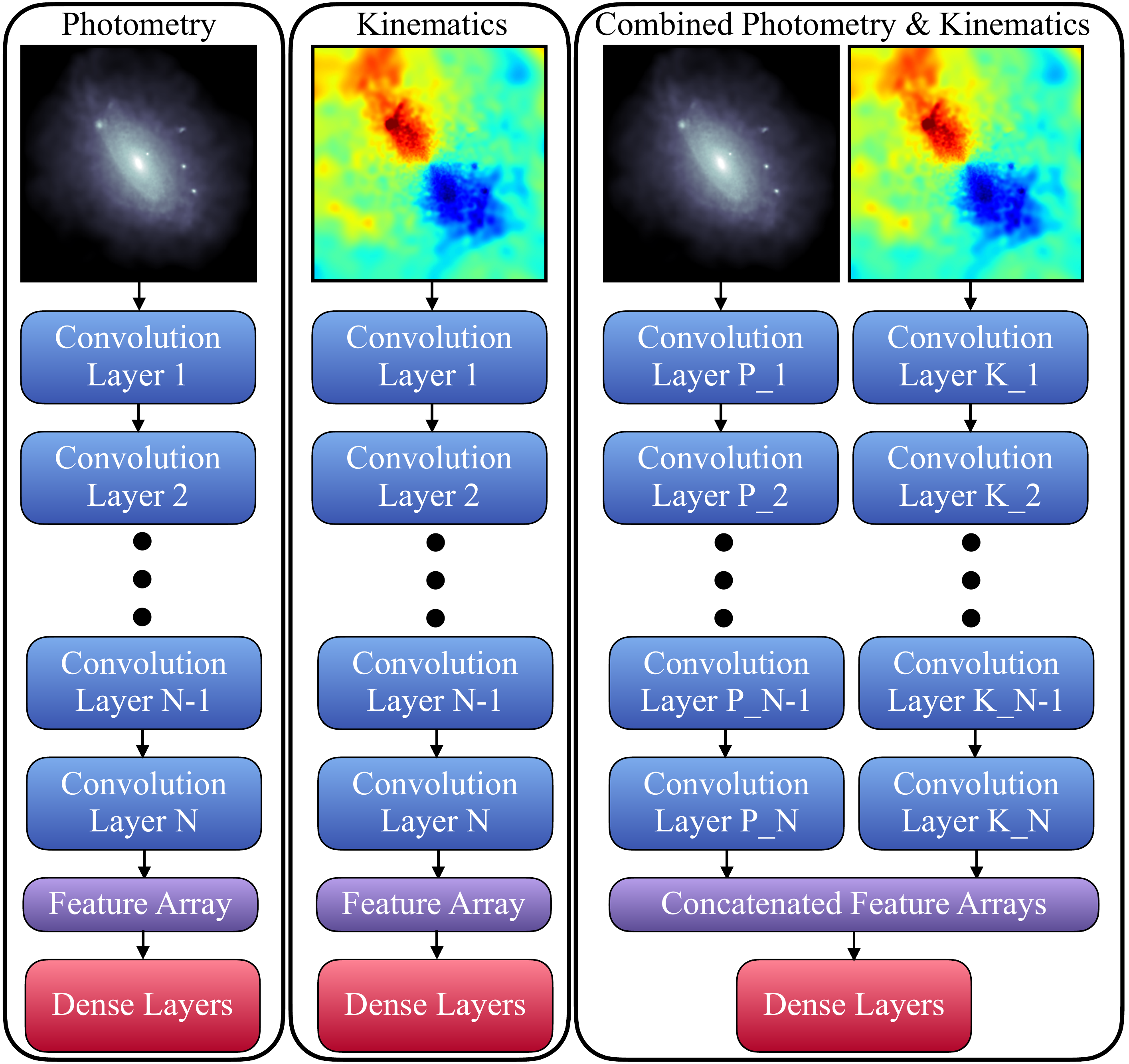}
   \caption[Classification Models]{Generalized convolutional neural network architectures used to classify TNG100 post-merger galaxies and examine the relative utility of photometric data, kinematic data, and combined photometry and kinematic data as input. In the case of AlexNet architectures, each convolution layer may comprise the filter (weight), activation, pooling, and batch normalization sub-layers. In the case of ResNet-V2 architectures, each convolution layer (technically, stage) comprises a set of residual blocks as shown in Table 1 in \cite{2015arXiv151203385H} -- which each contain multiple batch normalization, activation, and weight sub-layers as well as the skip connections across these blocks. For the single-input photometry and kinematic models, the features extracted from the final convolution layer are flattened and connected to the dense layers for classification. For the multi-input combined model, the feature arrays from the separate photometry and kinematic branches are concatenated for classification.}
    \label{fig:models}
\end{figure}

We use the stellar mass (hereafter, photometry) and velocity (hereafter, kinematics) moment maps to train and evaluate various flavours of convolutional neural networks (CNNs) for post-merger classification. CNNs are a class of deep learning models that are useful for pattern recognition in data which exhibit topological structure including series, images, and volumes \citep{lecun1995convolutional,lecun1998gradient,jarrett2009,10.1162/neco.2009.10-08-881,krizhevsky2010convolutional,2015Natur.521..436L}. 
Galaxies are randomly partitioned into training, validation, and test data sets \emph{by their IDs} using a $(70\%, 15\%, 15\%)$ split. As such, all four lines of sight for a given ID belong to the same partition, by construction. In addition, post-mergers and their controls always go into the same partition to ensure similarity between the mass, redshift, and environment distributions of post-mergers and controls within each partition. We took these careful measures to avoid possible memorization of features and subsequent over-fitting\footnote{In separate tests, however, we found that the models were insensitive to whether the data were partitioned based on galaxy ID vs line-of-sight and partitioning based on line-of-sight yielded no significant improvement.}. The final datasets for every model comprise $f_p \times 2,332 \times 2 \times 4 = f_p \times 18,656 = (13,056, 2,800, 2,800)$ images, where $f_p$ is the partition fraction for the training, validation, and test sets, respectively. Unless explicitly stated, the same training, validation, and test partitions are used for all models to guarantee fair comparison of model results. 

To investigate whether specific flavours of CNNs may facilitate improved performances with the imaging and/or kinematic data sets, we train and evaluate conventional AlexNet CNNs \citep{Krizhevsky:2012:ICD:2999134.2999257} and more sophisticated ResNet CNNs \citep{2015arXiv151203385H} with the improved pre-activation scheme proposed in \cite{2016arXiv160305027H} (ResNet-V2). Hyper-parameter options such as learning rates and schedulers, activation functions, batch and kernel sizes, and aspects of the architectures themselves (e.g. number of convolution and batch normalization layers) were comprehensively explored and results for every choice are available via our \href{http://www.comet.ml}{Comet.ML} project\footnote{\href{https://www.comet.ml/cbottrell/deepphokin?shareable=0Rkx8IFr6I5KR1UyVmReIq0Ej}{https://www.comet.ml/cbottrell/deepphokin}} -- which enabled detailed tracking and comparison of the performances of each architecture and set of hyper-parameters. All models were built, trained, and evaluated with \texttt{Keras} \citep{chollet2015keras} implemented in \texttt{Tensorflow 2}\footnote{\href{https://www.tensorflow.org}{https://www.tensorflow.org}} \citep{tensorflow2015-whitepaper}. To investigate the relative and combined roles of photometry and kinematics for post-merger classification, we optimize the models for three data sets: (1) photometry; (2); kinematics; and (3) photometry and kinematics combined. The moment maps were rebinned to $(128,128)$ for AlexNet architectures and $(224,224)$ for ResNet-V2 architectures. 

\subsubsection{AlexNet architectures}

From our hyper-parameter and architecture search, the best-fitting AlexNet models based on validation accuracy for data sets (1) and (2) were identical and used: batch size 32; five convolution layers with output depths $(32,64,64,128,128)$, convolution kernel sizes $(7,7,7,7,7)$ with stride 1, and $(2,2)$ max pooling after each convolution layer with stride 2; two fully-connected (dense) layers of size $(128,64)$ with $0.25$ dropout each \citep{srivastava2014dropout}; ReLU activation in both convolutional and fully connected layers \citep{nair2010rectified}; and a final layer dense layer comprising a single neuron with sigmoid activation. The models were optimized using a binary cross-entropy loss function and the Adadelta optimization algorithm \citep{2012arXiv1212.5701Z} with initial learning rate $0.1$. The learning rate was reduced by a factor of $0.1$ if the validation loss did not improve by at least $0.0001$ within a $10$-epoch patience. Early-stopping of training also used a $0.0001$ validation loss tolerance but $25$-epoch patience. An interesting result of our hyper-parameter exploration experiments is that models using ReLU activations in the convolution layers, which strictly produce positive values, provided the best accuracy for the velocity data set in which the data ranges from $-1$ to $1$. Other activation functions tested were $\tanh$ and Leaky ReLU \citep{maas2013rectifier}.

For the combined photometry and kinematic data model, we tested two broad architectures: multi-channel and multi-input. In the multi-channel case, the photometry and kinematics are provided to CNNs similar to the separate photometry and kinematics CNNs but as two-channel input, in the same way as multi-colour data is provided. In the multi-input case, the photometry and kinematics each have a separate feature extraction branch as illustrated in Figure \ref{fig:models}. The extracted features from each branch are then flattened and concatenated for the dense component. After experiments with both architectures, we settled on the multi-input model shown in Figure \ref{fig:models} which consistently yielded superior post-merger sample completeness and purity for both AlexNet and ResNet-V2 architectures. 

\begin{table}
\caption{Single-input ResNet-V2 architectures used in our classification analysis. The formula for the model architectures is \emph{identical} to those presented in \citet{2016arXiv160305027H} for models with more than 50 layers but typically with reduced numbers of residual blocks in stages Conv2-Conv5. The combined photometry and kinematics ResNet-V2 models comprise two branches, each including all blocks from Conv1 to Conv5. The feature arrays resulting from each branch are concatenated into a single feature array of length $4,096$ and connected to the dense layer as illustrated in Figure \ref{fig:models}.}
\label{tab:resnets}
\begin{tabular}{l  c  c  c}
\hline
Layer Name & Output Size & ResNet26-V2 & ResNet38-V2 \\
\hline
\hline
Conv1 & $112\times112$ & \multicolumn{2}{c}{$7\times7$, 64, Stride 2}  \\
\hline
& $56\times56$ &  \multicolumn{2}{c}{$3\times3$ Max Pool, Stride 2} \\
\hline
Conv2\_x &  $56\times56$ & $\begin{bmatrix} 1\times1,64 \\ 3\times3,64 \\ 1\times1,256 \end{bmatrix}\times2$ & $\begin{bmatrix} 1\times1,64 \\ 3\times3,64 \\ 1\times1,256 \end{bmatrix}\times3$  \\
\hline
Conv3\_x &  $28\times28$ & $\begin{bmatrix} 1\times1,128 \\ 3\times3,128 \\ 1\times1,512 \end{bmatrix}\times2$ & $\begin{bmatrix} 1\times1,128 \\ 3\times3,128 \\ 1\times1,512 \end{bmatrix}\times3$  \\
\hline
Conv4\_x &  $14\times14$ & $\begin{bmatrix} 1\times1,256 \\ 3\times3,256 \\ 1\times1, 1024 \end{bmatrix}\times2$ & $\begin{bmatrix} 1\times1,256 \\ 3\times3,256 \\ 1\times1,1024 \end{bmatrix}\times3$  \\
\hline
Conv5\_x &  $7\times7$ & $\begin{bmatrix} 1\times1,512 \\ 3\times3,512 \\ 1\times1, 2048 \end{bmatrix}\times2$ & $\begin{bmatrix} 1\times1,512 \\ 3\times3,512 \\ 1\times1,2048 \end{bmatrix}\times3$  \\
\hline
Feat. Array & 2048 & \multicolumn{2}{c}{2D Global Average Pool} \\
\hline
Dense & 1 & \multicolumn{2}{c}{1 Dense, Sigmoid}\\
\hline
\hline
\end{tabular}
\end{table}

\subsubsection{ResNet-V2 architectures}
The schematic CNNs illustrated in Figure \ref{fig:models} also generalize to the ResNet-V2 architectures. For ResNets, each convolution layer (technically, \emph{stages} for ResNets) in Figure \ref{fig:models} comprises a series of residual blocks which each necessarily contain multiple batch normalization, activation, and weight sub-layers as well as the skip connections across each block. We build 26-layer and 38-layer ResNet-V2 architectures which include $1\times1$ bottleneck convolution layers in each residual block for consistency with the deeper models presented in \cite{2016arXiv160305027H}. Table \ref{tab:resnets} outlines the ResNet26-V2 and ResNet38-V2 model architectures we built. Other than the number of residual blocks in stages Conv2-Conv5, all other aspects of the models (e.g. batch normalization layers, activations, shortcut connections) are \emph{identical} to the models presented in \citet{2016arXiv160305027H} with depths $\geq50$.

Our hyper-parameter search for the ResNet-V2 models was considerably smaller than for the AlexNet models. Nonetheless we searched over initial learning rates in $[0.005,0.01,0.05,0.1]$ and batch sizes $[16, 32, 64, 96, 128]$. The ResNet-V2 models had peak performance with batch sizes $32$ and initial learning rates $0.1$ for all data sets. Similar to the AlexNet models, improvement to the validation loss is monitored to reduce learning rates (by factors of $0.1$ to minimum $10^{-5}$) and to stop training with learning rate reduction and early-stopping patiences of 25 and 50, respectively. The combined photometry and kinematics models have separate feature extraction branches for each data set and feature arrays are concatenated ahead of a final, single-neuron dense layer with sigmoid activation.

Models were trained on single NVIDIA Tesla P100 (12 GB memory) or V100 (32 GB) GPUs. All AlexNet models were trained on the P100 GPUs with 6 cores along with the single-input ResNet26-V2 models. Multi-input ResNet26-V2 models and all ResNet38-V2 models were trained on V100 GPUs with 8 cores. Training times correspondingly varied. Typical training times for single(multi)-input AlexNet, ResNet26-V2, and ResNet38-V2 models were 40 mins (1 hour),  4 hours (7 hours), and 6 hours (10 hours), respectively.

\section{Results}\label{sec:results}

\begin{figure}
	\includegraphics[width=\linewidth]{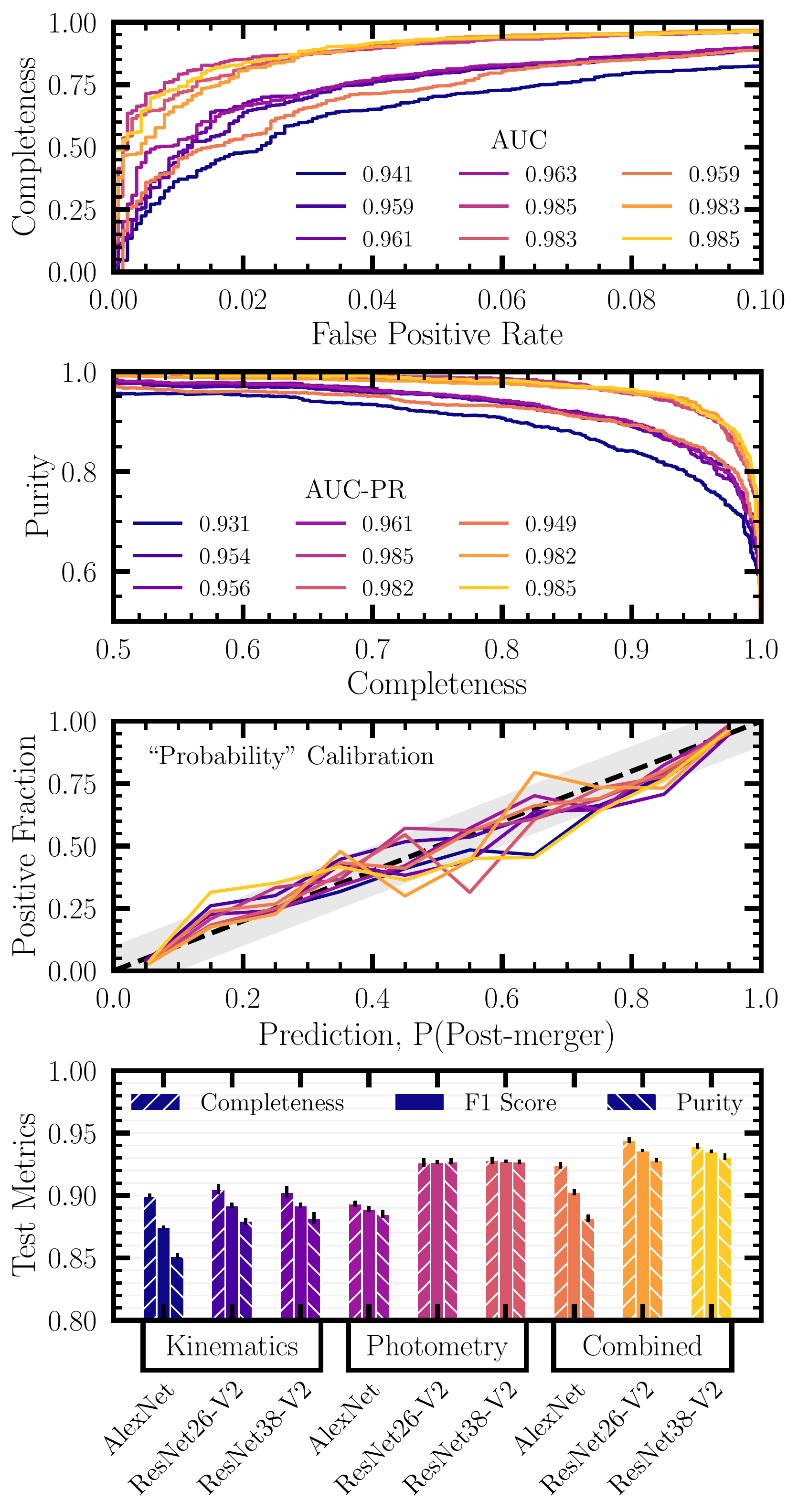}
   \caption[Model Metrics]{Evaluation metrics for our AlexNet and ResNet-V2 models on the kinematics, photometry, and combined data test samples. The coloured hatched and un-hatched bars in the bottom panel show the \emph{ensemble average} completenesses, $F_1$ scores, and purities computed over 11 runs with each model-data combination. Error-bars denote the standard deviation in the mean for each quantity. The coloured bars in the bottom panel provide the colour coding to the upper panels which show the completeness vs. false positive rate (top panel), purity vs. completeness (second panel from top), and the reliability diagnostic (probability calibration) curve (third panel from top) for the \emph{fiducial} runs (same random partition and weight initialization seeds). The top and second panels show the corresponding area-under-curve (AUC) scores and precision-recall AUC scores (AUC-PR) for each model and data set. The dashed line and gray shading in the reliability diagram show the one-to-one line and corresponding $\pm10\%$ offset.}
    \label{fig:metrics}
\end{figure}

Figure \ref{fig:metrics} shows the results of our best-fitting models on their corresponding test data. The lower panel of Figure \ref{fig:metrics} shows the \emph{ensemble average} completenesses, purities, and $F_1$ scores of each CNN architecture for each data set as colour-coded bars with upward, downward, and no hatch marks, respectively. Completeness is defined as the retention fraction of true post-mergers, $TP/(TP+FN)$, and the (balanced) purity is defined as the fraction of correctly classified post-mergers among all galaxies classified as post-mergers, $TP/(TP+FP)$, where $TP$, $FP$, and $FN$ are the numbers of true positives, false positives, and false negatives. $F_1$ is the harmonic mean of completeness and purity and is calculated as $F_1 = TP / ( TP + 0.5(FP+FN))$. The ensemble averages for each model-data combination are computed from the test results of $11$ models ($1$ fiducial and $10$ additional models) trained using the same model architecture and data type but different random data partitioning and model weight initialization seeds to capture stochasticity. The ensemble quantities shown in the bottom panel of Figure \ref{fig:metrics} are also tabulated in the upper half of Table \ref{tab:metrics} with corresponding standard errors for each average. We emphasize that \emph{these are estimates using a balanced test set}. Therefore, while the completenesses shown here are valid estimators of $P(+|\mathrm{Merger})$ from Figure \ref{fig:bayes}, the purities do not account for the natural imbalance of post-merger and non-postmerger galaxies and apply strictly to our balanced test data set. 

The bottom panel of Figure \ref{fig:metrics} provides the colour-coding of each model and data set to the lines in the upper three panels -- which show the results for a single \emph{fiducial} model for each model-data combination. The test data for each fiducial model (as well as training and validation data) comprises the same set of galaxies and all corresponding lines of sight for each of the photometry, kinematics, and combined data sets, as stated in Section \ref{sec:models}. We emphasize that \emph{the fiducial models are not necessarily the highest-performing models} from each ensemble for some metric. The fiducial models are simply models for which the partitioning of galaxies into training, validation, and test data was strictly identical -- such that all test galaxies have classification by every model type and every data set for comparison. The top panel of Figure \ref{fig:metrics} shows the relation between completenesses and false positive rates as well as the area-under-curve (AUC) score for each fiducial model and corresponding data set. The second panel shows the purities and completenesses and the corresponding purity/recall AUC scores. The third panel shows a reliability diagnostic for the merger probabilities -- which measures the fraction of post-mergers (positive fraction) in bins of the post-merger probabilities assigned by the model, $P(\mathrm{\text{Post-merger}})$. A model with well-calibrated probabilities will have a linearly declining fraction of post-mergers in each probability bin. The fiducial model metrics are tabulated in the lower half of Table \ref{tab:metrics}.

\begin{table}
\caption{Summary of the ensemble and fiducial AlexNet, ResNet26-V2, and ResNet38-V2 model metrics presented in Figure \ref{fig:metrics} for the photometric, kinematic, and combined data sets. The \emph{ensemble} mean and standard error in the mean for each quantity is computed over 11 runs with different data partition and model weight seeds to capture run-to-run and data partition stochasticity (bottom panel of Figure \ref{fig:metrics}). The \emph{fiducial} runs for each model-data combination all use the same random data partition and training seeds. The fiducial runs are used in subsequent analysis throughout this work. $^*$We again note that purities and F1 scores are sensitive to the prior for the true intrinsic merger fraction as shown in Figure \ref{fig:bayes} and explored thoroughly in \citep{2021MNRAS.504..372B}. The purities and F1 scores shown here are for balanced test data sets.}
\label{tab:metrics}
\setlength\tabcolsep{2.7pt} 
\begin{tabular}{l  l  c  c c}
\hline
& & Ensemble & Ensemble  & Ensemble  \\
Data Set & Model & $\langle$Completeness$\rangle$ & $\langle$F1 Score$\rangle^*$ & $\langle$Purity$\rangle^*$ \\
\hline
\multirow{3}{*}{Kinematics} & AlexNet & $89.9\pm0.2\%$ & $87.5\pm0.1\%$ &  $85.2\pm0.2\%$   \\
& ResNet26-V2 & $90.5\pm0.4\%$ & $89.2\pm0.2\%$ & $88.0\pm0.3\%$  \\
& ResNet38-V2 & $90.3\pm0.5\%$ & $89.2\pm0.2\%$ & $88.2\pm0.5\%$  \\
\hline
\multirow{3}{*}{Photometry} & AlexNet & $89.3\pm0.2\%$ & $88.9\pm0.2\%$ &  $88.5\pm0.4\%$   \\
& ResNet26-V2 & $92.6\pm0.4\%$ & $92.7\pm0.2\%$ &  $92.7\pm0.2\%$   \\
& ResNet38-V2 & $92.8\pm0.3\%$ & $92.8\pm0.1\%$ &  $92.7\pm0.2\%$   \\
\hline
\multirow{3}{*}{Combined} & AlexNet & $92.4\pm0.3\%$ & $90.2\pm0.2\%$ &  $88.2\pm0.3\%$   \\
& ResNet26-V2 & $94.4\pm0.2\%$ & $93.6\pm0.1\%$ &  $92.8\pm0.2\%$   \\
& ResNet38-V2 & $94.0\pm0.2\%$ & $93.5\pm0.2\%$ &  $93.1\pm0.3\%$   \\
\hline
& & Fiducial & Fiducial  & Fiducial  \\
Data Set & Model & Completeness & F1 Score$^*$ & Purity$^*$ \\
\hline
\multirow{3}{*}{Kinematics} & AlexNet & $90.1\%$ & $87.0\%$ &  $84.1\%$   \\
& ResNet26-V2 & $90.3\%$ & $89.5\%$ & $88.8\%$  \\
& ResNet38-V2 & $91.1\%$ & $89.7\%$ & $88.4\%$  \\
\hline
\multirow{3}{*}{Photometry} & AlexNet & $90.9\%$ & $90.1\%$ & $89.4\%$   \\
& ResNet26-V2 & $93.4\%$ & $93.4\%$ & $93.5\%$  \\
& ResNet38-V2 & $94.0\%$ & $93.5\%$ & $92.9\%$  \\
\hline
\multirow{3}{*}{Combined} & AlexNet & $91.9\%$  & $90.0\%$ & $88.3\%$   \\
& ResNet26-V2 & $95.0\%$ & $94.3\%$ & $93.5\%$  \\
& ResNet38-V2 & $94.5\%$ & $93.9\%$ & $93.4\%$  \\
\hline
\end{tabular}
\end{table}

\subsection{Sensitivity to the choice of CNN architecture}\label{sec:architectures}
We first consider differences between model architectures for a given data set -- taking the photometry data set as qualitatively representative of the other two datasets (middle three bars in the bottom panel of Figure \ref{fig:metrics}). Of the architectures considered, the AlexNet models performs most poorly on the photometry data set (as well as all other data sets). Its ensemble completeness, $F_1$ score, and purity are $(89.3\pm0.2, 88.9\pm0.2, 88.5\pm0.4)\%$ for the photometry data set. These results are comparable to those reported in \citet{2021MNRAS.504..372B} and \citet{2020ApJ...895..115F} with similar image-based data sets and AlexNet models. The deeper ResNet models perform markedly better on the photometric data by over $2\%$ in every metric -- indicating that the simpler AlexNet architectures are limiting the peak performance of merger classification CNNs used thus far in the literature.

A further point worth noting is that the AlexNet models used by \citet{2021MNRAS.504..372B} and \citet{2020ApJ...895..115F} were crucially trained and evaluated on synthetic images with survey-realism (e.g. \citealt{2019MNRAS.490.5390B}): careful insertion into high-quality images from the Canada France Imaging Survey \citep[CFIS]{2017ApJ...848..128I} in the case of \citet{2021MNRAS.504..372B} and the Cosmic Assembly Near-infrared Deep Extragalactic Legacy Survey \citep[CANDELS]{2011ApJS..197...35G,2011ApJS..197...36K} in the case of \citet{2020ApJ...895..115F}. Incorporating realistic observational features and properties (real skies, resolution considerations, processing artifacts, and neighbouring sources) into synthetic training images is \emph{essential} to a trained model's generalizability to other real/realistic data (\citealt{2019MNRAS.490.5390B}; see also \citealt{2021arXiv211100961C}). Our models are trained on idealized imaging and kinematic data and are therefore not suitable for application to data which contain these observational nuisances. However, because our models are trained and evaluated on synthetic data which \emph{do not include} realism, the evaluation metrics reported in Table \ref{tab:metrics} can be interpreted as \emph{upper limits} to the evaluation metrics for models of similar architecture that are trained and evaluated on synthetic data which \emph{do include} realism (i.e. the upper limits in which all observational limitations/nuisances could be perfectly removed from real/realistic data). Nonetheless, the performance metrics of \citet{2021MNRAS.504..372B} for survey-realistic synthetic CFIS images are similar to those presented here for idealized synthetic images using the same post-merger sample and control matching scheme. This similarity shows that our \emph{theoretical} upper limits can be closely approached by models that are trained and evaluated on realism-added images provided that the data are sufficiently high-quality\footnote{In contrast, \citet{2019MNRAS.490.5390B} found that models trained and evaluated on idealized synthetic images of the \citet{2019MNRAS.485.1320M} disc merger suite were markedly more accurate at characterizing mergers by phase than models trained and evaluated on the corresponding survey-realistic synthetic SDSS images ($96.0\%$ compared to $87.1\%$) -- which are considerably shallower and have poorer resolution than CFIS imaging.}. Lastly, the improved results obtained with our deeper ResNet models show that the model architecture may be a remaining factor in limiting the performances of the above-mentioned and other present merger classification models using more conventional AlexNet architectures.

The ensemble (fiducial) $F_1$ scores of the ResNet26-V2 and ResNet38-V2 models for the photometry data sets are $92.7\%\pm0.2\%$ ($93.4\%$) and $92.8\%\pm0.1\%$ ($93.5\%$), respectively and the corresponding completenesses and purities vary stochastically about these harmonic means. We interpret the consistency between these two depths as having reached converged performance with ResNet-V2 model architectures with this data set. The results for the photometry data set are qualitatively the same in the kinematics and combined data sets. Improved performances may still be gleaned by even more powerful models such as present state-of-the-art Vision Transformers \citep[ViTs]{2020arXiv201011929D,2021arXiv210604560Z}. In the sections that follow, we take the ResNet26-V2 architecture as our final, fiducial model architecture for analysis. 

\subsection{Utility of stellar kinematics and photometry}\label{sec:sensitivities}

\begin{figure*}
	\includegraphics[width=0.49\linewidth]{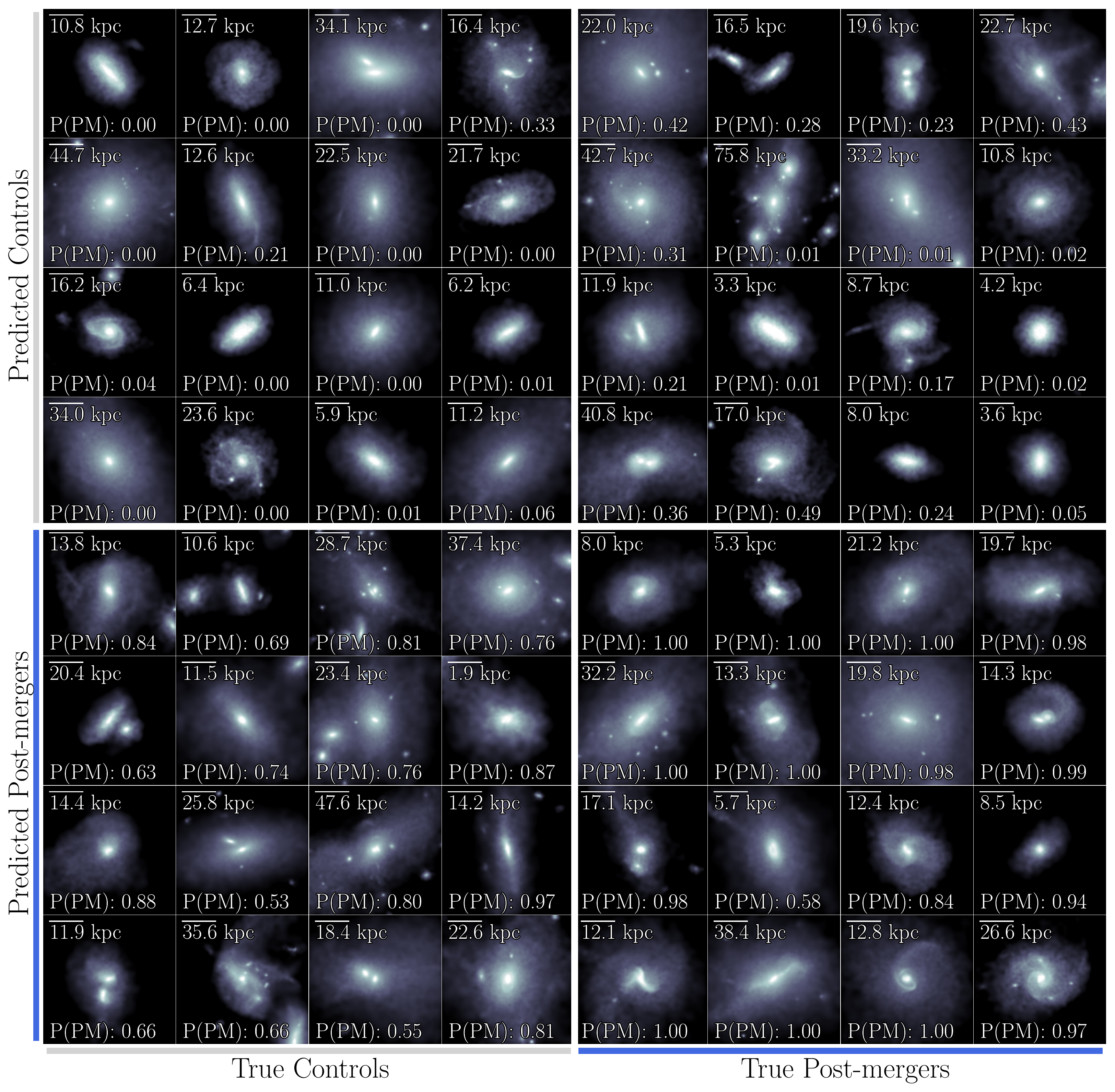}
	\includegraphics[width=0.49\linewidth]{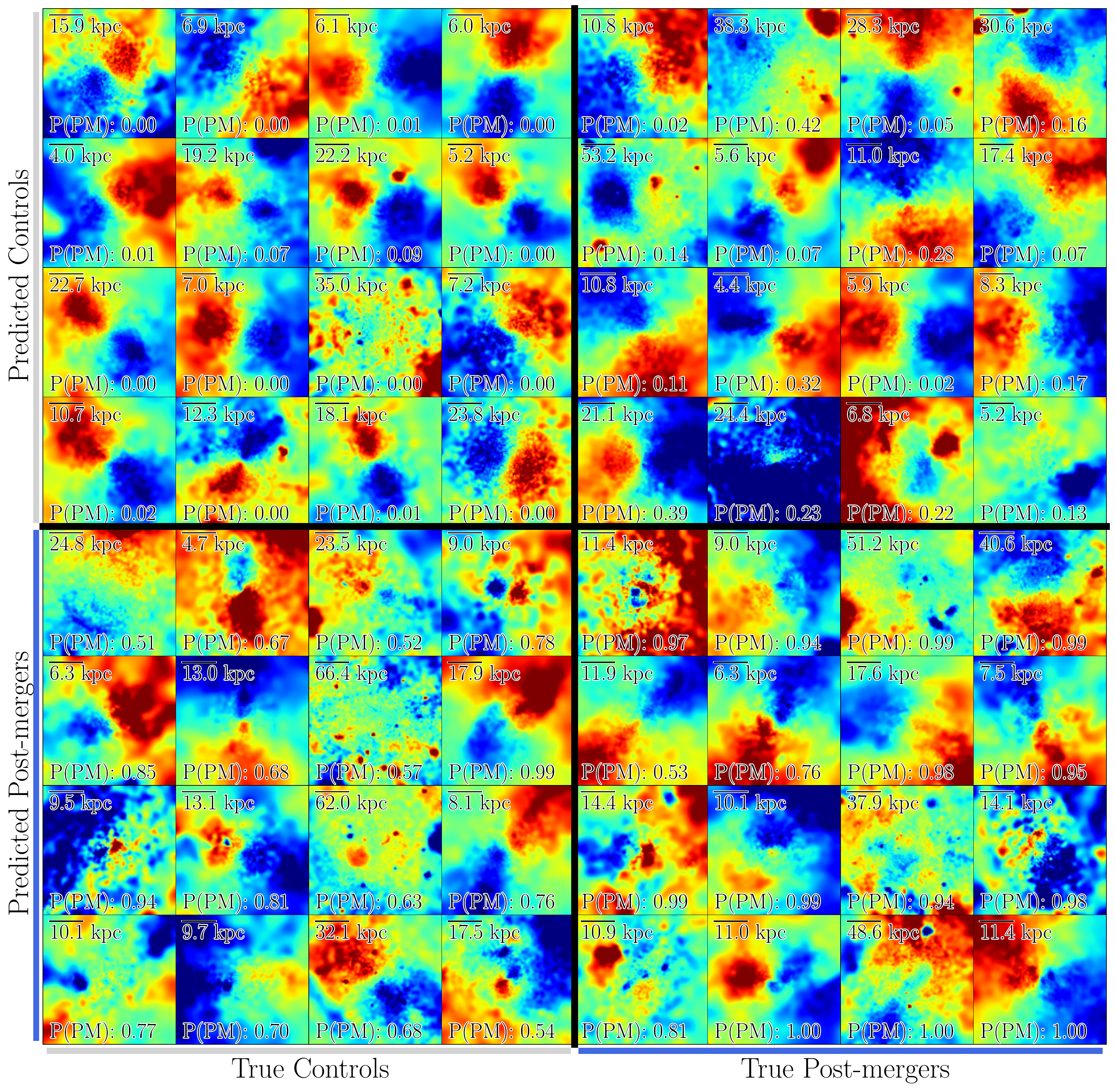}
   \caption[Photometry and kinematics mosaic]{Randomly selected photometry and kinematic maps for galaxies from the test data set whose outcomes were true negative (upper left quadrants in each panel), false negative (upper right quadrants), false positive (lower left quadrants), and true positive (lower right quadrants) when classified by the ResNet26-V2 photometry model (left panel) and kinematic model (right panel). Unlike Figure \ref{fig:mosaics}, the photometry and kinematic maps do not correspond to each other. Other examples showing the agreement/disagreement between photometric and kinematic models are shown in Figure \ref{fig:phokin_agreement}. \emph{On-line version only}: Each map's post-merger probability from the corresponding model is shown in the lower left.}
 \label{fig:pho_kin_mosaic}
\end{figure*}

\begin{figure*}
	\includegraphics[width=0.326\linewidth]{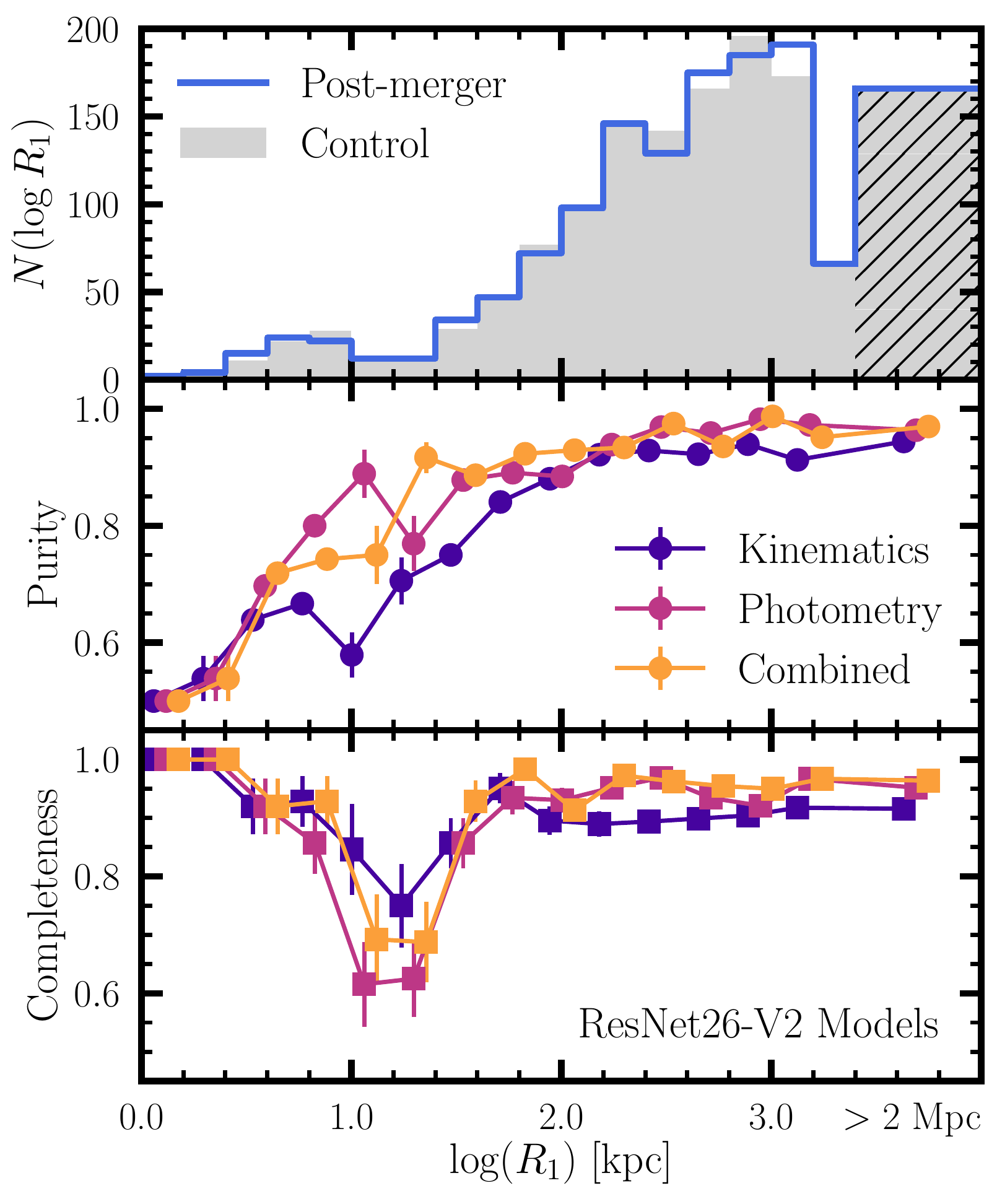}
	\hspace{-0pt}
	\includegraphics[width=0.33\linewidth]{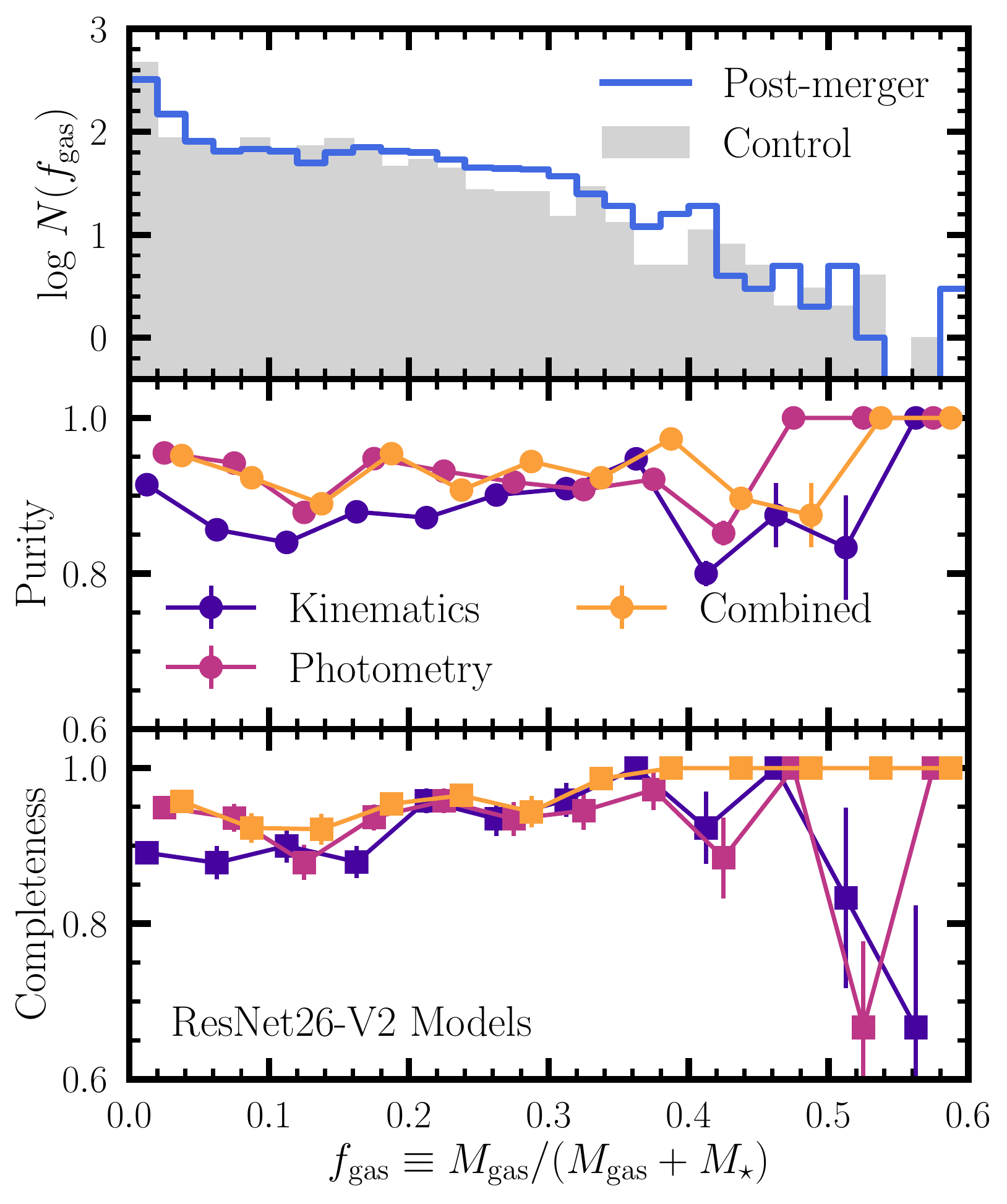}
	\hspace{-7pt}
	\includegraphics[width=0.329\linewidth]{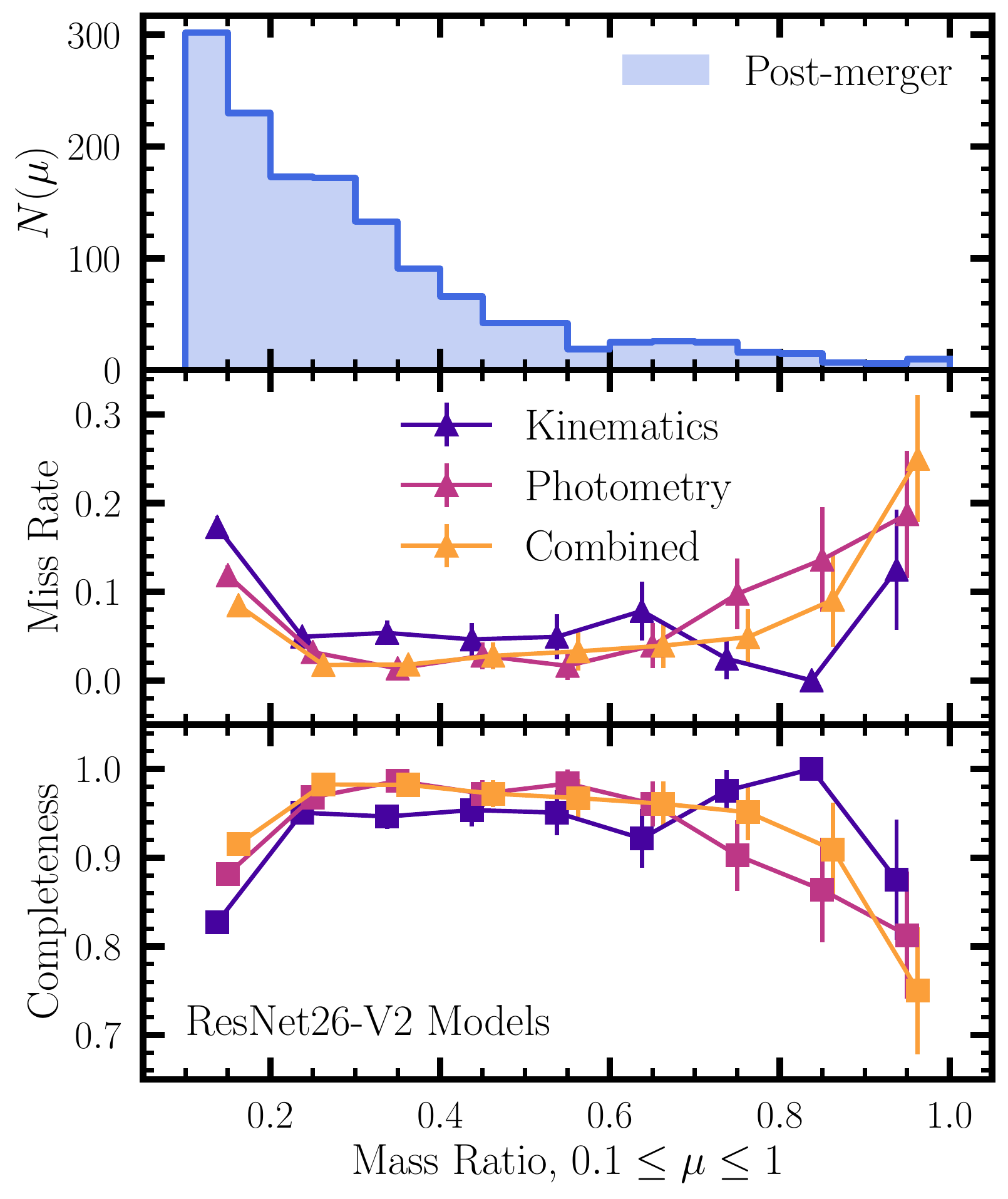}
   \caption[Sensitivities]{Sensitivities of post-merger classifications to 3D nearest neighbour distance (left), galaxy gas fraction (middle), and merger mass ratio (right). Histograms in the upper panels show the test-set distributions of post-mergers and controls in each parameter. The large hatched bin in the $R_1$ panel (middle) is for galaxies for which no $M_{\star,2} \geq 0.1 M_{\star,1}$ neighbour was found within 2 Mpc. Lower panels show the completenesses and purities of post-merger identification with the kinematic, photometric, and combined ResNet26-V2 models. For mass ratios, only post-mergers are shown and the purity panel is replaced with miss rate. Standard errors for the purities and completenesses in each bin are propagated from binomial classification statistics. The corresponding error bars are shown for all bins but are often smaller than the markers themselves.}
    \label{fig:sensitivities}
\end{figure*}

The ensemble metrics in the bottom panel of Figure \ref{fig:metrics} show that, when considered separately, \emph{stellar kinematic data have less overall utility for identifying post-merger galaxies than imaging data}. Although the third panel of Figure \ref{fig:metrics} shows that the probabilities from the fiducial kinematic models are as well-calibrated as the photometry models, all other panels show decreased performance metrics for models using the kinematic data compared to photometry (AUC, AUC-PR, completeness, $F_1$ score, purity). Each photometry model has an ensemble $F_1$ score that is at least $1.4\%$ higher than the corresponding model's score on the kinematics data set. In particular, the ensemble $F_1$ scores for the ResNet26-V2 and ResNet38-V2 models using photometric data are each 3.5\% higher than their counterparts using kinematic data. However, it is notable that these simple, normalized stellar velocity maps yield performances comparable to photometry -- particularly because the normalizations we employed do not allow the models to exploit the \emph{relative degree of rotation}. 

Figure \ref{fig:pho_kin_mosaic} shows photometry and kinematic mosaics for randomly-selected test set controls (four left columns) and post-mergers (four right columns) that are either classified as controls (four upper rows) or post-mergers (four lower rows) by the fiducial models. As such, the upper left, upper right, lower left, and lower right quadrants each show 16 randomly selected true negatives, false negatives, false positives, and true positives, respectively. In both the photometry and kinematic maps, the true negative samples (correctly classified controls) are visibly less disturbed than false positives, false negatives, and true positives, on average. The majority of the true negative kinematic maps exhibit ordered, axisymmetric rotation. Meanwhile, the true positives (correctly classified post-mergers) predominantly exhibit visually asymmetric morphologies and kinematics. However, the false positive and false negative maps in each panel reveal the problematic degree of visual overlap between the features spaces of controls and post-mergers. In the sections which follow, we explore this shared feature space and its sensitivity to 3D nearest neighbour distance, $R_1$, galaxy gas fraction, $f_{\mathrm{gas}} = M_{\mathrm{gas}}/(M_{\mathrm{gas}}+M_{\mathrm{\star}})$, and merger mass ratio, $\mu$ -- which specifically pertains to post-mergers (i.e. true positives and false negatives) in order to explore whether misclassifications occur in a certain parameter space for the photometry, kinematics, and combined-input models.

\subsubsection{Sensitivity to the proximity of the nearest companion}\label{sec:sens_r1}
The upper right and lower left quadrants of the photometry and kinematic panels in Figure \ref{fig:pho_kin_mosaic} show false negatives contributing to the incompletenesses of the models and false positives contributing to impurity, respectively. The false negatives contain many merger remnants which are themselves involved in secondary interactions. Similarly, the false positives contain many controls that are currently interacting with companions. In both cases, the classifications are confused by the on-going interactions with companions. In the case of the false positives, the interaction between controls and their companions yield morphological and kinematic features which overlap with post-mergers. Similarly, the interactions with companions for the false negatives yields reduced $P(\mathrm{PM})$ because of our enforcement of $R_1$ matching between post-mergers and controls -- which guarantees that companions should be as frequent in the control sample as the post-merger sample.

The left panel of Figure \ref{fig:sensitivities} quantitatively illustrates the role of companions in driving incompleteness and impurity in post-merger selection for the kinematics, photometry, and combined models. As described in Section \ref{sec:selection} on control matching, $R_1$ is the 3D distance between the target galaxy and the nearest neighbouring galaxy whose stellar mass is at least a tenth of that of the target. At $R_1\lesssim10$ kpc, the completenesses of all models are high and all true post-mergers are correctly classified. Meanwhile, the corresponding contamination due to the presence of companions at close proximity yields very low purities for all models: photometry, kinematics, and combined. The purity reaches $50\%$ in the lowest $R_1$ bin -- at which point the classification reliability for controls is no better than a coin-flip. However, the upper histogram of $R_1$ in the TNG100-1 post-merger and control sample shows that such cases are rare. The purity climbs rapidly with increasing $R_1$ in all models. These results, now considering stellar kinematics and combined models, are qualitatively consistent with the results of \citep{2021MNRAS.504..372B} who showed that the fraction of correctly classified controls climbs from $0\%$ at $R_1=2$ kpc to $\sim50\%$ at $10$ kpc and plateaus at $90\%$ around $100$ kpc.

Despite high completeness in $R_1<10$ kpc and beyond $40$ kpc, the left panel of Figure \ref{fig:sensitivities} shows that the models all exhibit dips in completeness specifically between $10-40$ kpc. The dips in this regime are commensurate with strong variations in the purities of the models as well as reduced galaxy counts. The variation between models are the result of trade-offs between completeness and purity. For example, the kinematic model's completeness has the least sensitivity to $R_1$ in this regime but its purity has the greatest sensitivity. On the other hand, the photometry model's completeness is most reduced but its purity is greatest. The challenge that is particular to $R_1<40$ kpc is that \emph{both} post-mergers \emph{and} controls include on-going interactions and fly-bys with companions in this regime. Observationally, there is a clear statistical enhancement of asymmetric morphological structure in galaxies with close physical companions compared to matched controls \emph{without companions} (e.g. \citealt{2007ApJ...666..212D,2010MNRAS.407.1514E,2013MNRAS.429.1051C,2016MNRAS.461.2589P}). In particular, \citet{2016MNRAS.461.2589P} specifically show that the enhancement in images is statistically significant at $r_p<40$ kpc, where $r_p$ is the \emph{projected} separation to the nearest spectroscopic companion. As $R_1$ decreases further (below 10 kpc), the purities continue to drop as controls increasingly resemble post-mergers, on average. Meanwhile, the completeness increases as all galaxies, control or post-merger, are indiscriminately given post-merger status by the models. In summary, in the pursuit of a pure post-merger sample, \emph{the addition of kinematic data is not helpful in removing contaminating companions from imaging data}.

\subsubsection{Sensitivity to the galaxy gas fractions}\label{sec:sens_gas}

Theoretical work has shown that the structures and kinematics of galaxy merger remnants and the longevity of merger-triggered features are sensitive to the merger gas content (e.g. \citealt{2005MNRAS.362..184B,2005A&A...438..507B,2006ApJ...650..791C,2006ApJ...645..986R,2010MNRAS.404..590L}). The middle panel of Figure \ref{fig:sensitivities} shows the sensitivity of model metrics to gas fraction. The completenesses and purities of the models are less sensitive to gas fractions than to $R_1$ but nevertheless increase steadily as gas fractions increase, on average. At $f_{\mathrm{gas}}>0.4$, model metrics become more sporadic, commensurate with small numbers of galaxy samples in this regime. However, galaxies with such high gas fractions are also expected to have irregular morphologies and disturbed kinematics that are unrelated to their merger histories (e.g. \citealt{2007ApJ...658..763E,2009Natur.457..451D,2011ApJ...730....4B}). This may also play a role in the confused behaviour of the models at high gas fractions. Nevertheless, the increasing trend for all models further establishes the sensitivity of merger-triggered features to the gas content in galaxies. The combined model predominantly tracks whichever of the photometry and kinematic does best in each gas fraction bin.

The photometry model has consistently higher purity than the stellar kinematic model for $f_{\mathrm{gas}}<0.3$. However, the completeness of the kinematic model tracks the photometric model very tightly for $f_{\mathrm{gas}}>0.1$. The key difference is at gas fractions $f_{\mathrm{gas}}<0.1$, where there is a $\sim5\%$ gap separating the two single-dataset models in a regime containing a very large fraction of the sample. Because our ``photometric'' images are simple surface density maps, our models cannot exploit global or local variations in colour or mass-to-light ratios from young stellar populations and active star-forming regions. As pointed out in \citet{2019MNRAS.490.5390B}, such insensitivity is most likely desirable -- as selection on colours or central starbursts limits the capacity in which resulting merger samples can be used in non-self-fulfilling experiments. In this study, however, we can only highlight that superior performance for photometry compared to kinematics is not due to exploitation of these higher-order correlations in the images.

\begin{figure}
	\includegraphics[width=\linewidth]{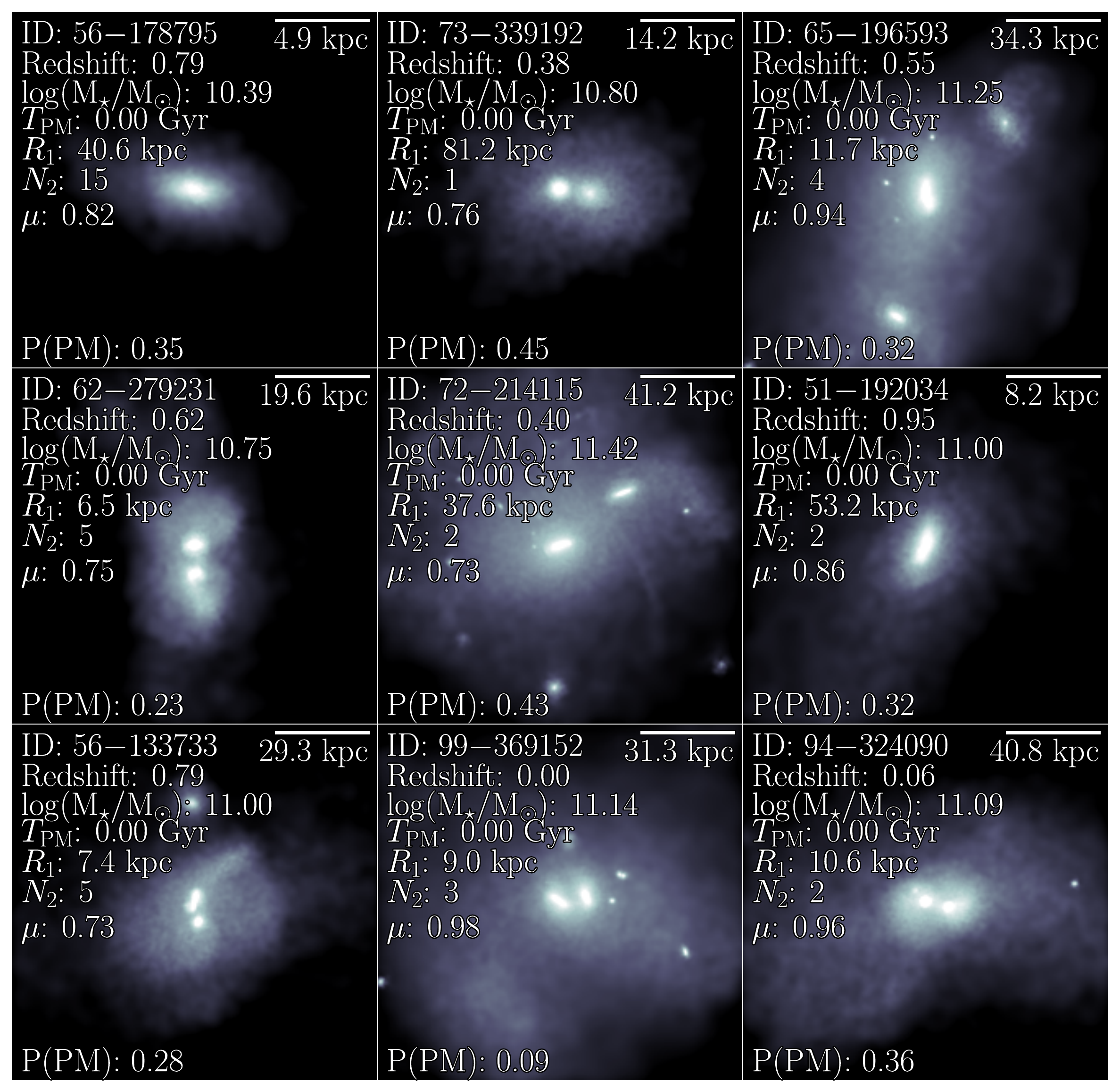}
   \caption[High- Mass Ratio False Negatives]{Randomly selected high- mass ratio galaxies from the post-merger test partition that were erroneously classified as controls by the ResNet26-V2 photometry model. These misclassifications contribute to the poor completeness at mass ratios $\mu\geq0.7$ in the right panel of Figure \ref{fig:sensitivities}. Their post-merger classification scores, $P(\mathrm{PM})$ are shown in the lower left of each panel. Each galaxy is exceptional in its own right. The majority shown are remnants which themselves are interacting with a third galaxy with $\mu\geq0.1$ as required by the $R_1$ parameter.}
 \label{fig:mr_fns}
\end{figure}

\subsubsection{Sensitivity to progenitor mass ratios}\label{sec:sens_mu}

The degree of morphological and dynamical transformation (including the longevity of transient features) triggered in galaxy merger remnants is sensitive to the mass ratio of the progenitors (e.g. \citealt{2003ApJ...597..893N,2005A&A...437...69B,2010MNRAS.404..575L}). The right panel of Figure \ref{fig:sensitivities} shows the sensitivity of post-merger classification metrics to the post-merger progenitor mass ratios. Only post-merger galaxies are considered here, so the purity is replaced with the miss rate -- which is is the fraction of true post-mergers erroneously classified as controls and therefore the complement to completeness. All models have poor completeness for $0.1\leq\mu<0.2$, our most minor mass ratio bin. The kinematic completeness is poorest compared to photometry in this regime. However, it is notable that the combined model yields improved completeness compared to photometry ($\sim4\%$). The kinematics therefore offer information that is complementary to photometry in this regime -- which alone cannot identify low-$\mu$ mergers with high completeness. 

All models have consistently higher completeness in the $0.2\leq\mu<0.6$ regime than their corresponding ensemble completenesses shown in Figure \ref{fig:metrics} and Table \ref{tab:metrics}. The photometry and combined models each average $97\%$ the kinematics average $94\%$ in this regime. This result illustrates how the performance of a merger classification model based on photometry or kinematics is intrinsically sensitive to the merger \emph{definition} -- in particular, whether minor mergers are considered. Models which focus on identifying major mergers (and therefore define the mergers as having $\mu>0.2$, for example) may see a large overall boost in the completeness with which such galaxies are retained. However, as illustrated in the upper panel, stricter definitions of mergers based on mass ratios will greatly reduce the merger sample size and, consequently, the prior for the intrinsic merger fraction, $P({\mathrm{Merger}})$, whose role in merger sample purity is shown in Figure \ref{fig:bayes}.

At higher mass ratios, $\mu\geq0.6$, the models decline in completeness -- with the exception of the kinematic model which temporarily rises in $0.6\leq\mu<0.9$ then declines in the highest mass ratio bin. This trend is unexpected given that the highest mass ratio mergers should have the most heavily disrupted remnants. The small number of calibration galaxies in this regime may be partially responsible for this behaviour. For example, there are only 3 erroneous classifications at $\mu\geq0.9$ out of a total of 16 galaxies for the photometry model. Ultimately, the decrease in completeness at high-$\mu$ with the fiducial ResNet26-V2 models is hard to reconcile without visual inspection. 

Figure \ref{fig:mr_fns} shows nine randomly selected examples of post-mergers erroneously classified as controls by the ResNet26-V2 photometry model with progenitor mass ratios $\mu>0.7$. Incidentally, all three of the $\mu>0.9$ misclassified post-mergers are shown. Each such galaxy (or more frequently, \emph{system}) is an exceptional circumstance. The majority of the images show ongoing mergers between a post-merger remnant and an additional galaxy. These interactions also satisfy the $\mu\geq0.1$ criterion required for the $R_1$ parameter to be allocated to the additional galaxy. The left panel of Figure \ref{fig:sensitivities} showed the high sensitivity of classification performance in the $0 \leq R_1\leq 40$ kpc range, in particular. Therefore, the decreased performances at high mass ratios are largely ascribed to a few exceptional circumstances that, when considering the small number of galaxies in this regime, each drive large changes in completeness. Using the ensembles of models trained with different partition and model initialization seeds, we confirm that the stochasticity in mass ratio completeness is generally high in the  $\mu\geq0.9$ regime and is not particular to any model architecture or data set.

\begin{figure*}
	\includegraphics[width=0.495\linewidth]{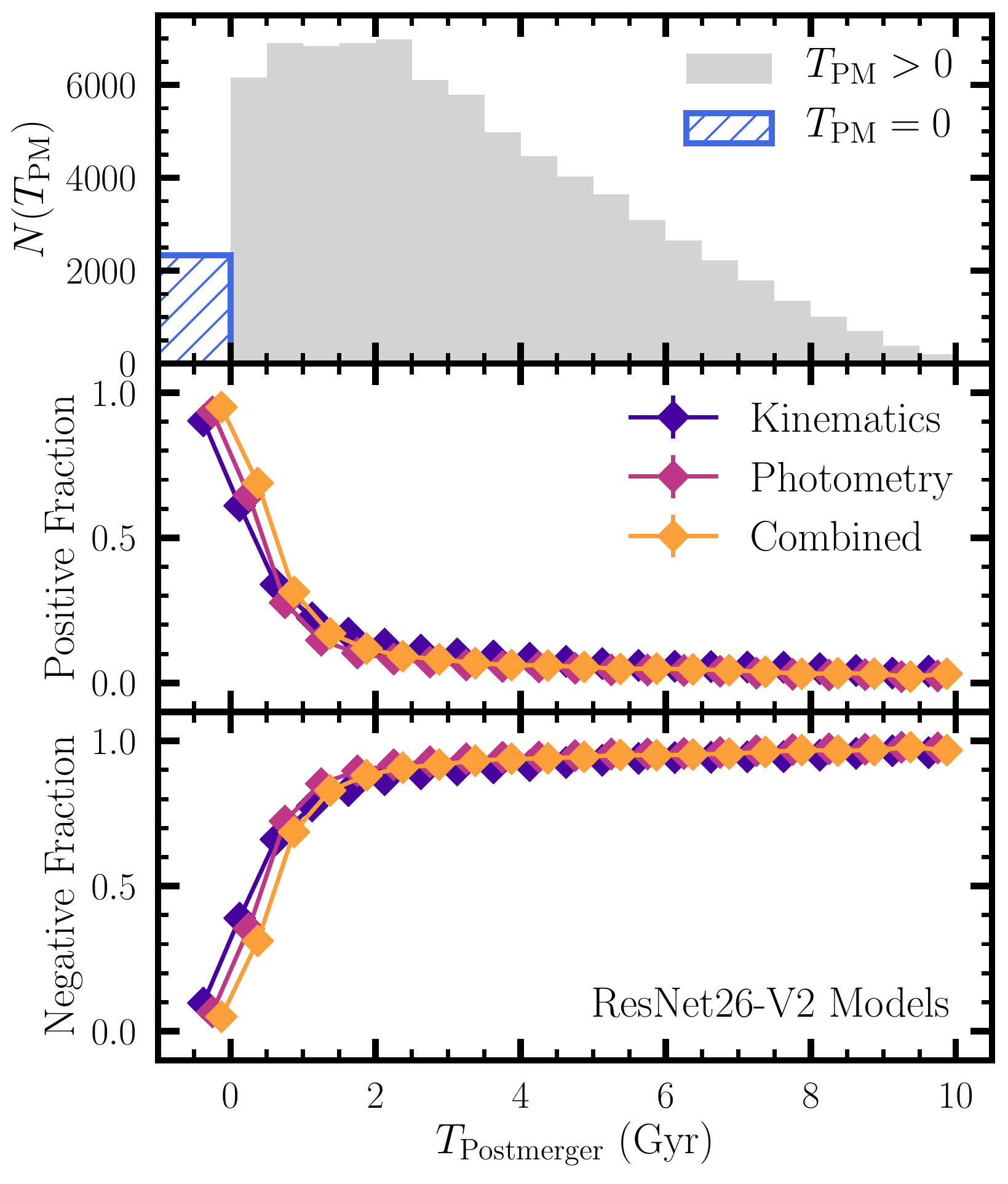}
	\includegraphics[width=0.495\linewidth]{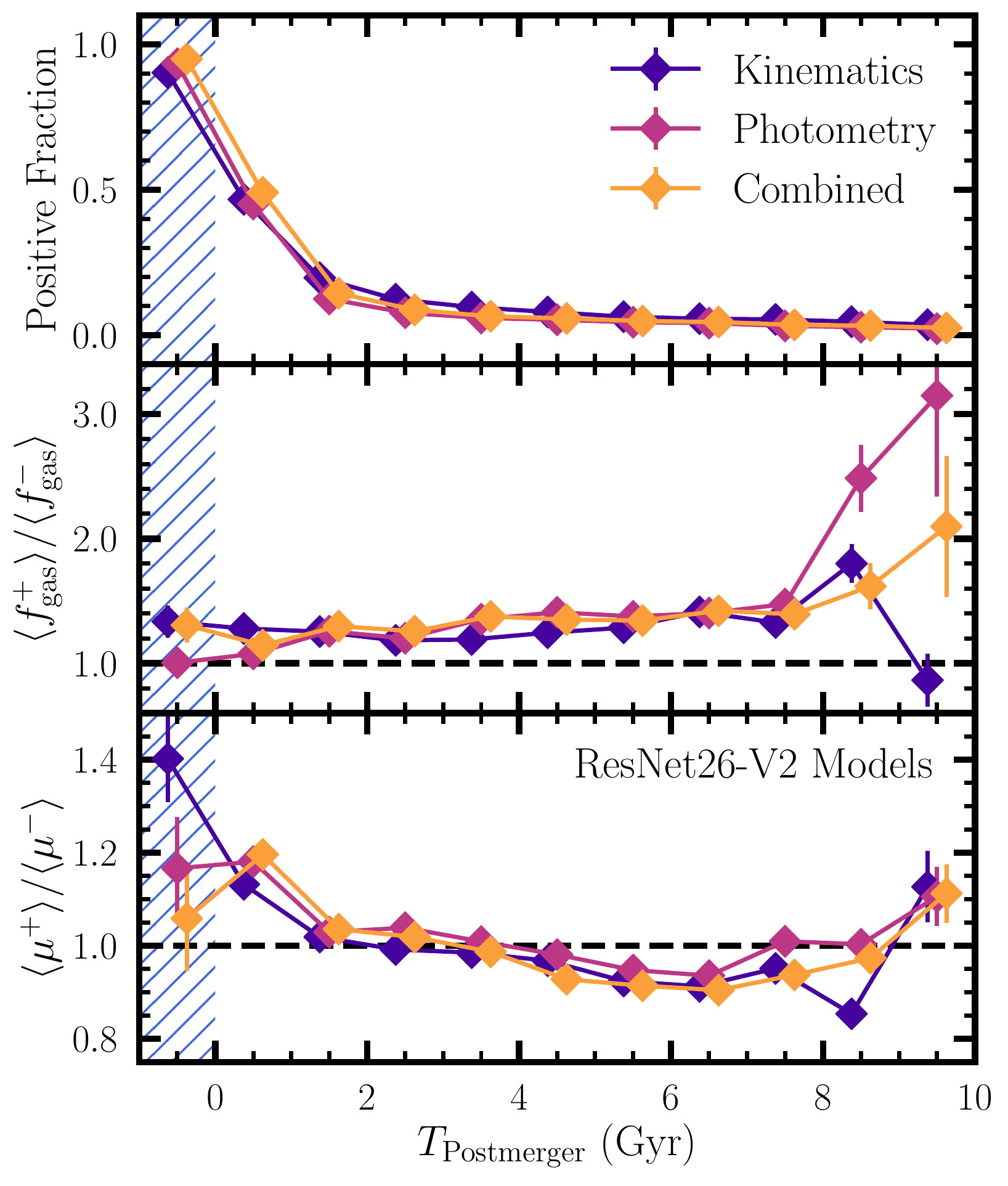}
   \caption[Sensitivity to $\tpost$]{Sensitivity of post-merger classification performance to the time since the most recent $\mu\geq0.1$ merger, $\tpost$. The left panels show the distribution in $\tpost$ for the full sample along with the completenesses (positive fractions) and miss rates (negative fractions) in bins of $\tpost$. The right panels show ratios of the ensemble averages in gas fraction, $\langle f^{+/-}_{\mathrm{gas}} \rangle$, and mass ratio, $\langle \mu^{+/-} \rangle$, for correctly ($+$) and incorrectly ($-$) classified post-mergers in each bin. The upper left panel shows all $78,584$ galaxies in the sample. Remaining panels consider only maps for the $78,584-85\%\times(2,332\times2)=74,620$ post-mergers and their descendants not used to train and validate the models.}
 \label{fig:tpost_class}
\end{figure*}

\subsection{Sensitivity to post-coalescence time, $\tpost$}\label{sec:tpost}

The \emph{observability timescale} for galaxy mergers and merger remnants is used to compute galaxy merger rates. Various formalisms for computing the galaxy merger rate exist (e.g. \citealt{2009MNRAS.394L..51B,2011ApJ...742..103L,2014MNRAS.445.1157C,2019ApJ...876..110D,2020ApJ...895..115F,2021arXiv210501675W}). Since no observational limitations are considered in this work, we do not estimate an ensemble observability timescale for post-merger galaxies -- which must be tuned to specific surveys due to the high sensitivity of merger identification to surface brightness and resolution conditions (e.g. \citealt{2008MNRAS.391.1137L,2019MNRAS.486..390B,2019MNRAS.490.5390B}). Instead, we examine the theoretical limits to which photometry and stellar kinematics can be used to distinguish post-mergers and controls in idealized data sets as a function of time since coalescence, $T_{\mathrm{PM}}$.

To investigate the sensitivity of model performance to $\tpost$, we use the same best-fitting models described in the previous sections. These models are applied to a separate test data set comprising \emph{all galaxies} from TNG100-1 that satisfy the same cuts for post-mergers outlined in Section \ref{sec:selection} but no restrictions on $\tpost$. The resulting sample comprises a total of $78,584$ galaxies with stellar masses $\logMstar\geq10$ and redshifts $z\leq1$ that are the descendants of a merger whose progenitors have stellar mass ratios $\mu>0.1$. As such, all $78,584$ galaxies are now considered merger remnants for the purpose of assessing classification performance sensitivity to the time since the merger, $\tpost$. As before, the stellar kinematics and photometry of each such galaxy is observed in four line-of-sight orientations for a total of $314,336$ photometric and kinematic maps, each. 

\subsubsection{Completeness as a function of post-coalescence time}\label{sec:tpost_comp}

The upper left panel of Figure \ref{fig:tpost_class} shows the distribution of $\tpost$ for the all 78,584 galaxies in the data set. However, the $2,332 \times 2 \times (70\%+15\%)= 1,982$ post-merger and matched control galaxies which belonged to the models' training and validation partitions are omitted from the performance analysis in all other panels. \emph{Immediate} post-mergers with $\tpost=0$ (test galaxies from previous sections) are allocated to $\tpost\leq0$ bins in Figure \ref{fig:tpost_class}. The middle-left panel shows the ensemble completenesses (positive fraction) of ``post-merger'' galaxies in bins of $\tpost$ for the kinematic, photometry, and combined models. The lower-left panel shows the miss-rate (negative fraction).

Completeness drops rapidly with $\tpost$. By $\tpost=0.5$ Gyr, the $\tpost=0$ Gyr completeness drop by $\sim30\%$ with no changes in advantage between the photometry or kinematic models. The interpretation of this result is that these $30\%$ of merger-remnants are no longer distinguishable from control galaxies whose $\tpost>2$ Gyr by definition. By $\tpost=1$ Gyr, the model completenesses drop to $\sim30\%$ and the stellar kinematic data become most useful. The slightly higher utility of stellar kinematic data persists for the remaining range in $\tpost$, albeit with very limited overall completeness. This improved utility of stellar kinematics at high $\tpost$ is qualitatively consistent with the results of \citet{2016ApJ...816...99H} and \citet{2021ApJ...912...45N} in disc galaxy merger simulations with observational realism. However, the \emph{degree} of improved utility is here shown to be very limited when considering a representative galaxy population -- even with idealized data. We discuss the roles of observational limitations and sample generalizability in Sections \ref{sec:realism} and \ref{sec:samples}.

By $\tpost=2$ Gyr, the completenesses drop to $\sim10\%$. Since the control sample used to calibrate the model has $\tpost>2$ Gyr by definition, all galaxies with $\tpost>2$ Gyr should technically fall into the control category in the eyes of the models. However, some galaxies are classified as post-mergers all the way to $\tpost=10$ Gyr. Statistically, the origin of the apparent persistence of post-merger features out to high $\tpost$ is due to the non-zero false-positive-rates of the models, $P(+|\mathrm{\text{Non-post-merger}})$, which is the basis for impurity. The false-postitive-rates for the ResNet26-V2 kinematic, photometric, and combined models were $(11.4\%, 6.5\%, 6.5\%)$, respectively. Therefore, while some post-merger galaxies may exhibit features which distinguish them from controls to high $\tpost$, the contribution of contamination in this range from galaxies with irregular morphologies and kinematics whose origins are unrelated to the merger is statistically dominant. 

\subsubsection{Preferences on mass ratios and gas fractions}\label{sec:tpost_prefs}

Certain merger scenarios may be more likely to produce persistent morphological and stellar kinematic features. Two merger parameters which may be related to the longevity of merger-triggered features are the gas fractions and mass ratios of the mergers. The right panels of Figure \ref{fig:tpost_class} show the ratios of the ensemble average gas fractions, $\langle f_{\mathrm{gas}} \rangle$, and mass ratios, $\langle \mu \rangle$, of correctly ($+$) and incorrectly ($-$) classified post-merger galaxies in each $\tpost$ bin (true positives and false negatives, respectively). The lower right panel shows that the stellar kinematic model is more likely to assign post-merger status to $\tpost=0$ Gyr galaxies originating from high mass ratio progenitors. The photometric model is less sensitive to mass ratio and the combined model even less so -- reflecting results shown in Figure \ref{fig:sensitivities}. All models exhibit similar sensitivity to mass ratio in the $0<\tpost\leq1$ Gyr range. Afterward, positively and negatively classified galaxies exhibit no ensemble differences in mass ratio and differences between models are small apart from noisy behaviour at $\tpost>8$ Gyr. 

As in Section \ref{sec:sensitivities}, the gas fractions here are the \emph{current} gas fractions. This choice was made so that two effects could be explored: (1) the potential persistence of features due to high gas fractions at low-$\tpost$ and (2) potential biases for galaxies with high gas fractions and high-$\tpost$\footnote{A possible concern with this choice is that high $\tpost$ merger descendants \emph{might} have reduced gas fractions relative to their progenitors due to accelerated conversion of gas into stars in the post-merger stage or as a result of feedback-driven outflows (e.g. \citealt{2008MNRAS.384..386C,2020MNRAS.493.3716H,2020A&A...635A.197D,2021MNRAS.504.1888Q}). However, using TNG100, \citet{2021MNRAS.504.1888Q} show that quenching via either of these mechanisms amongst star-forming post-merger galaxies is rare (about $\sim5\%$ of merger descendants they consider) -- only a factor of 2 more frequently than matched star-forming control galaxies. Several observational studies also find higher atomic \citep{2015MNRAS.448..221E,2018MNRAS.478.3447E} and molecular \citep{2018MNRAS.476.2591V,2018ApJ...868..132P} gas fractions among post-merger galaxies -- commensurate with other theoretical results (e.g. \citealt{2019MNRAS.485.1320M,2019MNRAS.490.2139R}).}. The middle-right panel of Figure \ref{fig:tpost_class} shows that the ensemble of post-mergers correctly classified by their \emph{stellar} kinematics have $34\%$ higher gas fractions than post-mergers incorrectly classified as controls at $\tpost=0$ Gyr. This bias is manifested in the poor completeness and purity of the stellar kinematic model with respect to photometry at gas fractions $f_{\mathrm{gas}}<0.3$ in Figure \ref{fig:sensitivities}. This bias in the stellar kinematic is broadly uniform to high $\tpost$. Meanwhile, the photometric model initially exhibits no preference on gas fraction, but has a steadily increasing preference with increasing $\tpost$. High-$\tpost$ galaxies have quiescent merger histories (apart from $\mu<0.1$ mergers which are neglected). Among galaxies whose $\tpost$ are large, an increased preference for galaxies with high gas fractions is commensurate with increased incidence of asymmetric and/or clumpy structures observed in galaxies with high- gas fractions in the local Universe and at high-redshift (e.g. \citealt{2007ApJ...658..763E,2010ApJ...710..979O,2015ApJ...807..134G,2015ApJ...800...39G,2019ApJ...874...59S}) -- which are easily visually confused with merger-triggered morphologies. 

\section{Discussion}\label{sec:discussion}

In this section, we discuss caveats and consider our results in the context of other works assessing the utility of stellar kinematic data for merger/post-merger identification. 

\subsection{The effect of observational limitations on the performance gap between photometry and kinematics}\label{sec:realism}

We have considered the theoretical utility of photometry, kinematics, and combined data for identifying post-merger galaxies using \emph{idealized} data sets. Specifically, we have shown that the idealized stellar kinematic information is less useful in predicting post-merger status than idealized photometry and only rarely offers advantages when combined with photometry. The gap between the performances of kinematics and photometry is expected to be greatly broadened when considering instrumental limitations -- which disproportionately degrades the stellar kinematic information relative to photometry. State-of-the-art highly multiplexed IFS surveys providing stellar kinematics for $\sim10^3$ to $10^4$ galaxies such as MaNGA \citealt{2015ApJ...798....7B}, CALIFA \citep{2012A&A...538A...8S}, SAMI \citep{2012MNRAS.421..872C}, and future IFS surveys such as HECTOR  \citep{2018SPIE10702E..1HB} have effective spatial resolution $\sim2''-2.5''$ after reconstructing observations from individual fibres. Additional limitations compared to our idealized map comparison are the observational footprints (fields of view, FoV) of the IFUs used in these surveys -- which are limited to $\sim1-3\; R_e$ compared to the $\sim10\; R_e$ we consider\footnote{Technically our FoVs are equal to 10 half-stellar-mass radii. So this translation is only exact by assuming that all stellar particles have mass-to-light ratios $M/L=1$}, where $R_e$ is the radius enclosing $50\%$ of the optical light in a photometric bandpass. 

McElroy et al. (in prep) specifically examine the role of field-of-view in distinguishing isolated, pair, and post-merger galaxies using line-of-sight stellar kinematics derived a suite of 27 binary merger simulations covering a large range of orbital configurations at fixed mass ratio, $\mu=0.4$ \citep{2019MNRAS.485.1320M}. McElroy et al. (in prep) degrade and crop maps to fields-of-view commensurate with the SAMI and HECTOR fibre-based IFUs and the Multi-Unit Spectroscopic Explorer \citep[MUSE]{2010SPIE.7735E..08B} observational footprints at $z=0.4$ -- roughly the median redshift of the SAMI survey galaxy sample \citep{10.1093/mnras/stu2635}. Kinematic asymmetries are extracted from the resulting maps using \textsc{Kinemetry}\footnote{\href{http://www.davor.krajnovic.org/software/index.html}{http://www.davor.krajnovic.org/software/index.html}} \citep{2006MNRAS.366..787K}. They specifically show that the utility of the stellar kinematic data degrades significantly with the size of the observational footprint -- obtaining $98\%$, $82\%$, and $50\%$ completenesses for post-mergers in MUSE, HECTOR, and SAMI fields of view, respectively.

Meanwhile, top-tier wide-field optical imaging surveys such as CFIS \citep{2017ApJ...848..128I} and the Hyper Suprime-Cam Subaru Strategic Program \citep[HSC-SSP]{2018PASJ...70S...4A} have (1) galaxy counts that outnumber IFS surveys by orders of magnitude and (2) sufficient depth and spatial resolution that they negligibly degrade post-merger classification performance in the local Universe out to at least $z\lesssim0.3$ \citep{2021MNRAS.504..372B}. Indeed, our AlexNet results with idealized photometry do not differ greatly from those of \cite{2021MNRAS.504..372B} -- who used a similar CNN and the same post-merger sample inserted into real CFIS tiles out to $z=0.3$ using an extension of the \realsim{} survey-realism code tailored to the CFIS survey \citep{2017ApJ...848..128I}. The commonality between the results from \cite{2021MNRAS.504..372B} in the survey-real images and our own with idealized photometry shows that there is no particular performance degradation that arises from observational considerations in such high-quality images (such as sky brightness, atmospheric blurring, and crowding by neighbouring sources in projection). 

In short, the already poorer performance of the idealized kinematic models with respect to photometry is only going to be degraded further when the spatial resolution, spectral resolution, and footprint sizes of IFS instruments are factored in. In other words, current and forthcoming IFS surveys are not going to the be the magic bullet for identifying mergers. 

\subsection{Reconciling our results with models calibrated on disc galaxy merger simulations}
\label{sec:samples}

While our results for idealized synthetic TNG100 galaxies show that the kinematic data has limited utility for identifying post-mergers compared to images, conflicting results are demonstrated in analyses using realistic synthetic imaging and kinematic maps for disc merger simulations -- which show that the kinematics can play a much more important role in the identification of recent post-mergers.

\citet{2021ApJ...912...45N} go to great lengths to make highly realistic synthetic stellar kinematic and imaging observations using a separate suite of five binary merger simulations covering a range of mass ratios on a fixed orbit \citep{2018MNRAS.478.3056B}. Specifically, \citet{2021ApJ...912...45N} produce MaNGA-like maps that include dusty kinematic radiative transfer from \texttt{sunrise}: \citealt{2006MNRAS.372....2J,2010MNRAS.403...17J} \emph{and} stellar LOSVD inference from corresponding synthetic spectra. From these realistic maps, they show that estimators derived from the stellar kinematic data (including both velocity and velocity dispersion maps) were \emph{generally} less useful than non-parametric morphological parameters derived from images for the two more minor mergers they consider, $\mu=(0.1,0.2)$, in agreement with our results. However, for their three major mergers ($\mu\geq0.33$ in their definition), \citet{2021ApJ...912...45N} find that omitting the stellar kinematic information from their photometry+kinematic combined analysis suppresses their accuracies, completenesses, and F1-scores of their models by $20-50\%$ -- meaning that some these kinematic parameters may indeed be the dominant predictors for major merger status. This contrasts with our results where we find that photometry works just as well by itself as when the kinematics are included.

The apparent conflict between our results and \citet{2021ApJ...912...45N} may be due to differences in the merger and control samples used in our analyses. Specifically, \citet{2021ApJ...912...45N} highlight that the importance of kinematic data for major merger detection may be artificially inflated in their analysis by the scope of scenarios covered in their merger suite. The suite of simulations used to calibrate their models are mergers between disc-dominated morphologies (e.g. as in \citealt{2008MNRAS.391.1137L,2016ApJ...816...99H,2019ApJ...872...76N,2019MNRAS.490.5390B}; McElroy et al. in prep). In each of these studies, transformation of dynamically cool, disc-dominated progenitors to lower-spin, dynamically heated remnants with more prominent stellar bulges is \emph{generally} expected in most merger configurations \citep{1977egsp.conf..401T,1983MNRAS.205.1009N,2008ApJS..175..356H,2013ApJ...778...61T,2014MNRAS.444.3357N,2018MNRAS.480.2266M}. However, dispersion-supported stellar bulges and elliptical morphologies are dynamically stable configurations and not unique to recent merger remnants (see \citealt{1997ARA&A..35..637W,2008gady.book.....B,2016ASSL..418..161F,2016ASSL..418..263G} for reviews). The consequence is that morphological and dynamical properties can be exploited to distinguish progenitor and remnant galaxies within disc merger suites that do not transfer to merger identification in a heterogeneous population of galaxies and merger scenarios. 

In particular, \citet{2021ApJ...912...45N} found that the most useful kinematic parameters for distinguishing mergers and non-mergers in their MaNGA synthetic kinematic maps track the spin-down of angular momentum and the growth of a higher- velocity dispersion component ($\lambda_{R_e}$ \citealt{2007MNRAS.379..401E}, and the mean, variance, and kurtosis of the velocity dispersion maps). We consider this result to be analogous to the results from \citet{2019MNRAS.490.5390B} that also showed artificially little morphological confusion between remnants and progenitors of 23 disc galaxy mergers with their CNNs -- even in images that included detailed realism. While we do not incorporate any observational effects in our maps in this work, we draw our galaxy and merger-remnant populations from a morphologically and dynamically heterogeneous and diverse galaxy sample \citep{2019MNRAS.483.4140R,2019MNRAS.487.5416T,2019MNRAS.489.1859H,2020arXiv200204182D}. Therefore, our analysis provides a theoretical performance limit to the identification of post-merger galaxies using photometry and kinematics \emph{with and for} a representative galaxy population (albeit, allowing for potential improvements from more sophisticated classification models). This difference rationalizes the more minor performance improvement when incorporating kinematic information into our models. 

\subsection{Outlook for merger detection in representative and heterogenous galaxy samples}\label{sec:outlook}

\begin{figure}
	\includegraphics[width=\linewidth]{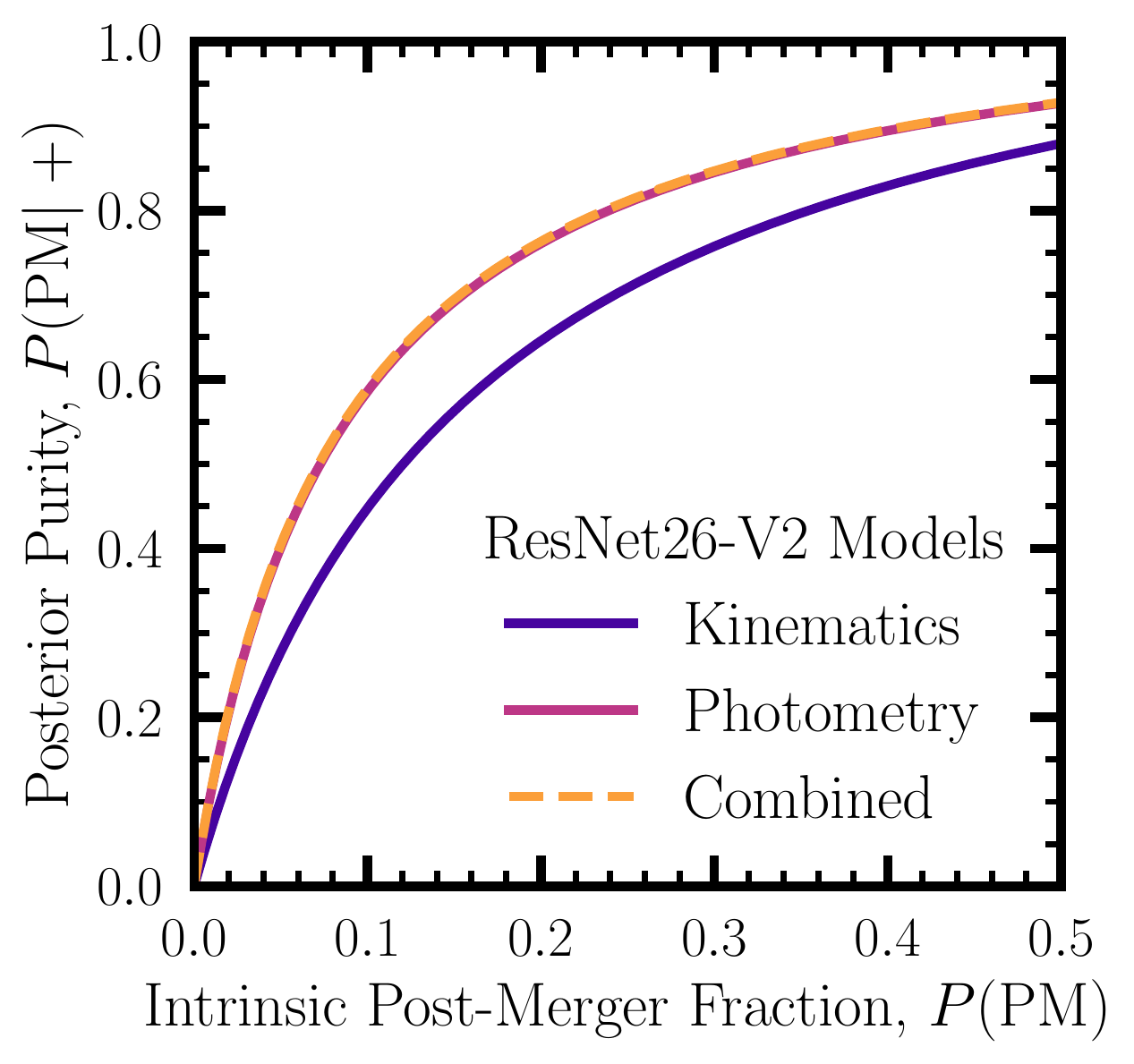}
   \caption[Posterior Purities]{Expectations for the posterior purities of post-merger samples, $P($PM$|+)$, that would be obtained with the fiducial kinematic, photometric, and combined ResNet26-V2 models for a range of priors for the intrinsic post-merger fractions, $P$(PM), using Equation \ref{eq:bayes}. For reference, at $P$(PM)$=0.5$, the posterior purities are the same as those shown in Table \ref{tab:metrics} for these models. The expected posterior purity declines rapidly with decreasing intrinsic post-merger fraction. The purities expected from the combined model do not improve upon those obtained from photometry alone.}
 \label{fig:purities}
\end{figure}

We have shown that even \emph{idealized} stellar kinematic data has limited utility in improving merger selection in representative galaxy samples. We have also shown that deep, highly non-linear models such as the CNNs employed here can be used to obtain completenesses in excess of $90\%$ for post-merger classification. However, the major hurdle to obtaining pure and complete merger samples from a heterogenous galaxy population is the purity -- particularly considering the intrinsic class imbalance between post-mergers and non-post-mergers in the real Universe as illustrated in Figure \ref{fig:purities}. Coloured lines in Figure \ref{fig:purities} show the expected posterior purities of post-merger samples obtained by our fiducial photometry, kinematic, and combined models, $P($PM$|+)$, for a range of intrinsic (true) post-merger fractions, $P($PM$)$. The photometry and combined models almost perfectly overlap -- showing that incorporating stellar kinematic information does not improve sample purities compared to photometry data alone. However, the posterior purities of all models decline quickly with decreasing intrinsic post-merger fractions. Even our highest-performance photometry model, with $P(+|\mathrm{PM})=93.4\%$ completeness and $P(-|\mathrm{\text{Non-PM}})=93.5\%$ specificity, could only be expected to glean a post-merger sample that is $P(\mathrm{PM}|+) = (12.7, 26.9, 43.1, 61.5, 82.7, 93.5)\%$ pure 
using priors $P(\mathrm{PM})=(1.0, 2.5, 5.0, 10, 25, 50)\%$, respectively. A new approach is required. In particular, in a follow-up to this work, we develop and examine a novel approach which may circumvent the role of these priors.

The remaining issue when training on a simulated data set is the domain change between simulations and observations -- despite efforts to bridge this gap (e.g. \citealt{2017MNRAS.467.2879B,2017MNRAS.467.1033B,2019MNRAS.486.3702S,2019MNRAS.489.1859H,2019MNRAS.490.5390B,2019ApJ...872...76N,2020ApJ...895..115F,2021ApJ...912...45N,2021MNRAS.504..372B,2021arXiv211100961C}). However, the remaining questions for the domain change is no longer whether observational effects can be properly accounted for but whether the intrinsic feature space of galaxies is consistent. In particularly promising investigation, \citet{2021MNRAS.501.4359Z} use autoregressive deep learning models to quantitatively examine the intrinsic differences in the feature spaces of TNG galaxies and galaxies from the Sloan Digital Sky Survey (SDSS) Legacy Images \citep{2009ApJS..182..543A}. Specifically, \citet{2021MNRAS.501.4359Z} show there is a high degree of similarity between the observed morphologies of galaxies in the SDSS and the TNG100 and TNG50 simulations, but also show tensions -- particularly in the dearth of quiescent, small and/or compact galaxies in the simulations. Meanwhile, some tensions exist in the stellar kinematic relations of observed and simulated galaxies (e.g. \citealt{2019MNRAS.484..869V}). So far, it is unclear the role these differences play when using synthetic observations of galaxies from simulations to calibrate merger classification and characterization models. Nevertheless, the similarities in the detailed structures and dynamics of galaxies in simulations and observations is compelling.

\section{Conclusions}\label{sec:conclusions}

We have characterized the theoretical degree to which galaxy merger remnants (post-mergers) can be distinguished from a heterogenous population of galaxies using idealized synthetic photometry and stellar kinematic data from the TNG100 cosmological hydrodynamical simulation (Section \ref{sec:moments}, Figure \ref{fig:mosaics}). We employ several convolutional neural network (CNN) architectures of increasing depth and sophistication to classify post-mergers and non-postmergers using the synthetic data sets (Section \ref{sec:models}, Table \ref{tab:resnets}, Figure \ref{fig:models}). In particular, we examined the relative and combined performances of photometry and stellar kinematic using single- and multi-input CNNs, respectively, in obtaining complete and pure post-merger samples. We were particularly motivated by whether stellar kinematic data may offer a complementary basis to photometry for post-merger galaxy identification -- thus enabling greater post-merger sample purity in overlapping wide-field imaging and highly-multiplexed integral field spectroscopy surveys. Our main results are as follows:
\begin{enumerate}
\item \textbf{The AlexNet or ``vanilla'' CNN architecture used ubiquitously in the merger and tidal feature identification literature has poorer performance metrics compared to more sophisticated models such as the ResNet class of CNNs} (Section \ref{sec:architectures}, Table \ref{tab:metrics}, Figure \ref{fig:metrics}). This reduction in performance is observed in each of the photometry ($3.8\%$ decrease in ensemble average F1 score), kinematic ($1.7\%$), and combined ($3.4\%$) data sets. These apparently incremental increases in performance are nonetheless important when considering the role of priors in post-merger sample purities (Figure \ref{fig:bayes}). We show that our ResNet-V2 CNNs have converged results and that increasing depth yields no further improvement in model metrics. However, even more sophisticated CNNs and exciting new models such as Vision Transformers may yet yield even greater performances. 

\item \textbf{The utility of stellar kinematics for identifying post-merger galaxies is limited relative to photometry} (Section \ref{sec:sensitivities}, Table \ref{tab:metrics}, Figure \ref{fig:metrics}). Using the same model architectures, the photometric model ensemble average F1 scores, the geometric mean of completeness and purity (balanced), are $1.4\%$ (AlexNet), $3.5\%$ (ResNet26-V2), and $3.6\%$ (ResNet38-V2) \emph{higher} than for the corresponding kinematic models. In particular, we highlight that the \emph{idealized} stellar kinematic data considered here do not incorporate observational limitations such as observational footprint, atmospheric and instrumental resolution, or line-of-sight-velocity distribution extraction from spectra. Such considerations, which would disproportionately affect the stellar kinematic data compared to imaging, should be expected to further broaden the gap in performance between imaging and kinematics (Section \ref{sec:realism}).

\item \textbf{The utility of stellar kinematics when used \emph{in tandem} with photometry (combined models) is limited} (Section \ref{sec:sensitivities}, Table \ref{tab:metrics}, Figure \ref{fig:metrics}). We show that incorporating the idealized stellar kinematic data in a dual-input model yields changes in ensemble average F1 scores by $3.1\%$ (AlexNet), $0.9\%$ (ResNet26-V2), and $0.7\%$ (ResNet38-V2) compared to photometry. Here again, we discuss the expectations for the disproportionate role of observational limitations on observed stellar kinematic data quality in Section \ref{sec:realism} and argue that current and forthcoming IFS surveys will not be the magic bullet for merger identification when combined with high-quality imaging surveys. We reconcile the less significant role of stellar kinematics in identifying merger remnants in a heterogenous galaxy population with the more significant role inferred from disc galaxy merger simulations in Section \ref{sec:samples}.

\item \textbf{All models exhibit high sensitivity to the proximity of a physical companion} (as opposed to projected companions; Section \ref{sec:sens_r1}, Figure \ref{fig:pho_kin_mosaic}, left panel of Figure \ref{fig:sensitivities}). For mass ratio $\mu>0.1$ companions at separations $R_1\lesssim40$ kpc, all models exhibit large fluctuations in their completenesses and purities -- commensurate with the difficulty in distinguishing post-merger remnants from on-going mergers and interactions. The models show more minor sensitivities to gas fractions and the mass ratio range we consider (middle and right panels of Figure \ref{fig:sensitivities}).

\item \textbf{Classification performance is highly sensitive to the time since coalescence, $\tpost$, for post-merger galaxies} (Section \ref{sec:tpost_comp}, Figure \ref{fig:tpost_class}). For example, by $\tpost=(0.5, 1, 2)$ Gyr, our highest-performance photometry model drops from $93.4\%$ completeness at $\tpost=0$ Gyr to $(64, 21, 10)\%$, respectively. All models show post-merger selection preferences on higher- mass ratio merger progenitor scenarios for $0\leq\tpost\lesssim2$ Gyr which disappears at later times. The models also preferentially give post-merger status to galaxies with high gas fractions (Section \ref{sec:tpost_prefs}, Figure \ref{fig:tpost_class}).

\end{enumerate}

Overall, we find that kinematic information is not expected to improve upon the performances of models which use imaging from deep, wide-field photometric surveys. The prior from the small expected fraction of merging galaxies in the local Universe (based on spectroscopic galaxy pair studies) presents outstanding challenges to obtaining pure and complete merger samples. Our work calls for new approaches to merger and post-merger identification which will overcome or mitigates the role of this prior and contamination in merger samples -- one of which we investigate in a forthcoming study.

\section*{Acknowledgements}
CB gratefully acknowledges support from the Natural Sciences and Engineering Research Council of Canada (NSERC) as part of their post-doctoral fellowship program [PDF-546234-2020]. MHH gratefully acknowledges support from the William and Caroline Herschel Postdoctoral Fellowship fund. DRP and SLE gratefully acknowledge NSERC for Discovery Grants which helped to fund this research. This research was enabled by computational resources provided by Compute Canada on the Cedar cluster (\href{www.computecanada.ca}{www.computecanada.ca}). This research made use of \texttt{Astropy}\footnote{\href{http://www.astropy.org}{https://www.tensorflow.org}}, a community-developed core Python package for Astronomy \citep{astropy:2013, 2018AJ....156..123A}. Kavli IPMU is supported by World Premier International Research Center Initiative (WPI), MEXT, Japan.

\section*{Data Availability}
The synthetic observations used in this paper are available on reasonable request to CB and MH. The TNG100 simulations (and now all simulations in the TNG suite) are publicly available at \href{https://www.tng-project.org}{https://www.tng-project.org/}. The \href{https://github.com/cbottrell}{GitHub} page for this project, including code and documentation, and \href{http://www.comet.ml}{Comet.ML} workspace will become publicly accessible upon completion of our follow-up to this study.



\bibliographystyle{mnras}
\bibliography{References} 




\appendix

\section{Agreement between photometric and kinematic models}

\begin{figure*}
	\includegraphics[width=0.45\linewidth]{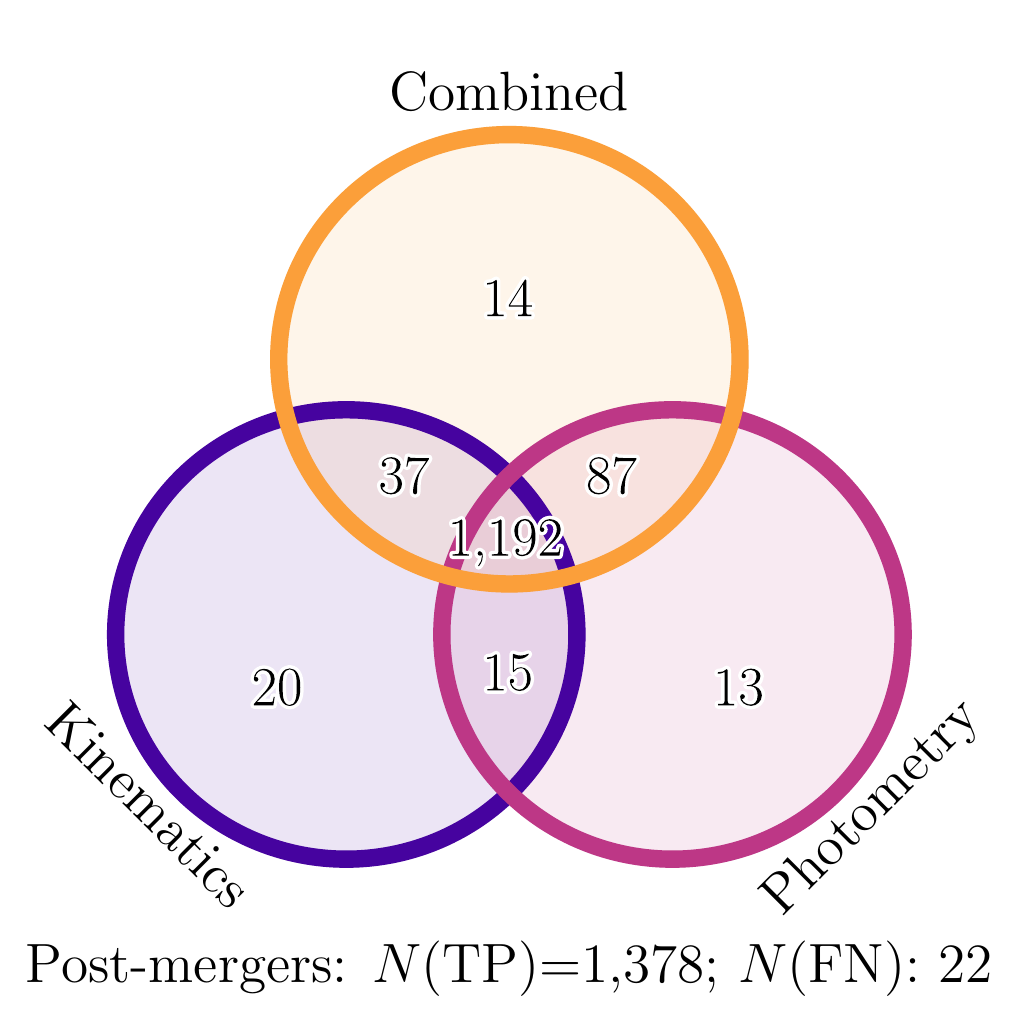}
	\includegraphics[width=0.45\linewidth]{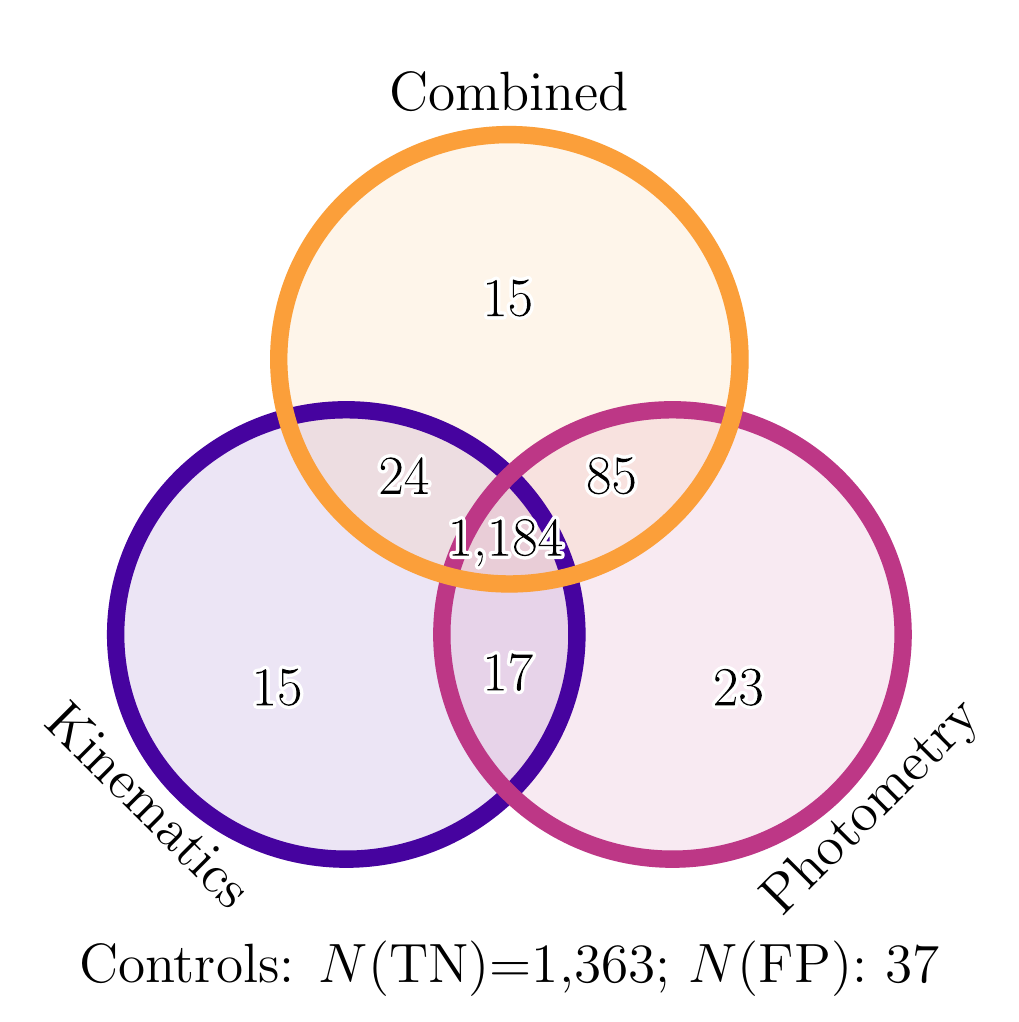}
   \caption[Venn diagrams of classification agreement]{Venn diagrams for the agreement between the photometric, kinematic, and combined ResNet26-V2 models for the post-merger (left) and control (right) test samples. The total numbers of post-mergers and controls in the test set are $1,400$ each.}
 \label{fig:venn_diagrams}
\end{figure*}

\begin{figure*}
	\includegraphics[width=\linewidth]{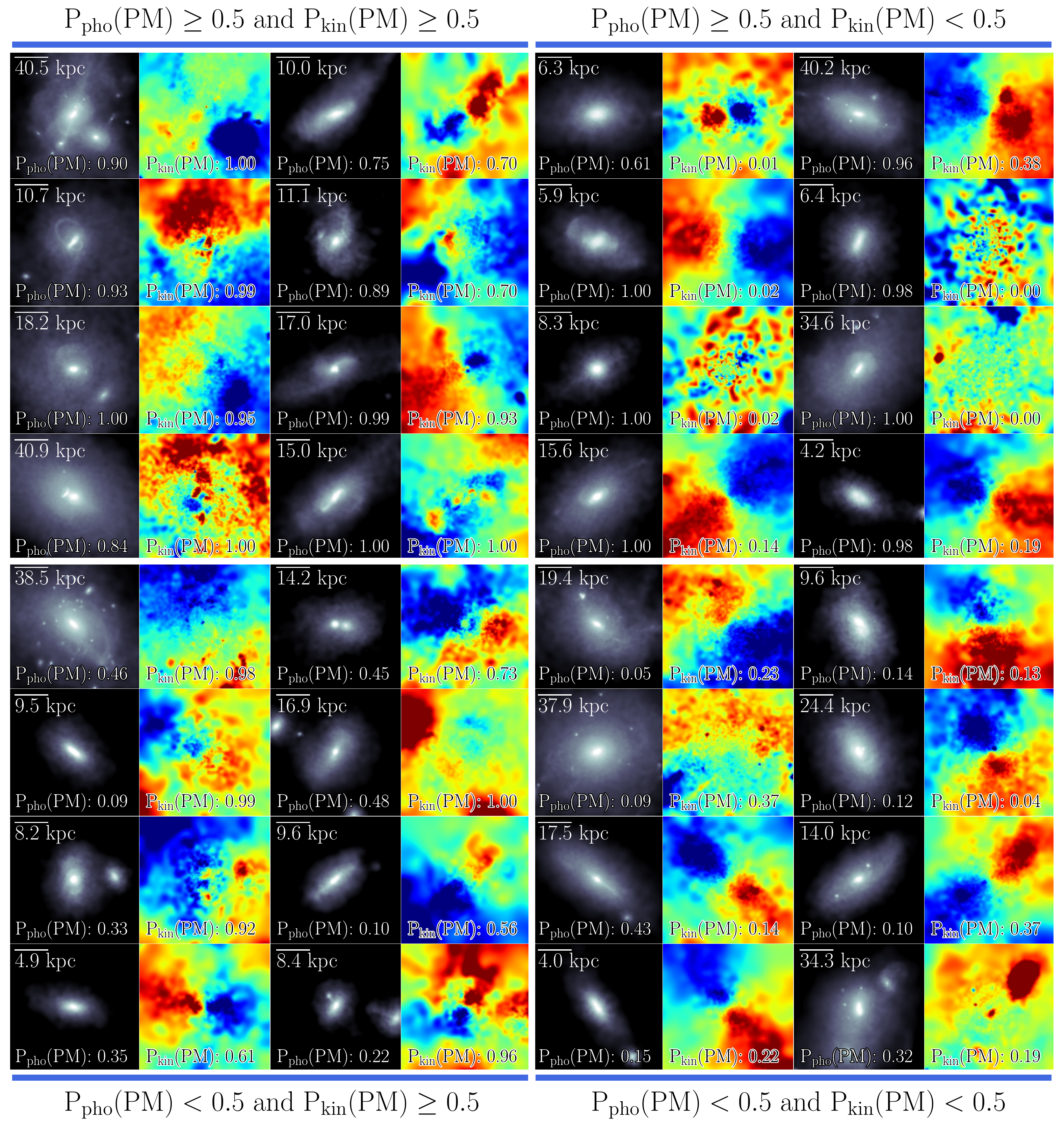}
   \caption[Agreement between kinematics and photometry]{Examples of post-merger galaxies for which the photometric and stellar kinematic ResNet26-V2 model classifications agree and disagree. All galaxies considered are post-mergers. The four quadrants show the photometry and kinematic maps for eight galaxies upon which the photometric and kinematic models: (upper left) correctly agree that the galaxy is a post-merger; (lower right) incorrectly agree that the galaxy is a control; (upper right) the kinematic model incorrectly disagrees with photometric model; and (lower left) the photometric model incorrectly disagrees with the kinematic model. }
 \label{fig:phokin_agreement}
\end{figure*}

Figure \ref{fig:mosaics} showed mosaics of post-mergers and controls classified by the photometric and kinematic models. We now consider cases where these model agree (correctly and incorrectly) or disagree in their classifications, specifically for post-merger galaxies. Figure \ref{fig:venn_diagrams} shows Venn diagrams for the agreement and disagreement between the photometry, kinematic, and combined model classifications for both post-mergers (left) and controls (right) in the test set. The number of galaxies correctly classified by at least one model, $N$(TP) and $N$(TN), and number of galaxies missed by all models, $N(FN)$ and $N$(FP), are shown below each panel. While the large majority of galaxies are correctly classified by all models (see also Table \ref{tab:metrics} and Figure \ref{tab:metrics}), several galaxies yield disagreements. 

There is a small and roughly similar number of galaxies for which only one model correctly classifies the galaxy (non-intersecting sectors in Figure \ref{fig:venn_diagrams}). Excluding the intersection of all models (central sector of each panel), the most highly-populated sectors of both the post-merger and control Venn diagrams are where the photometric and combined models both correctly classify a galaxy, but the kinematic model incorrectly disagrees ($87$ and $85$ for post-mergers and controls, respectively). In contrast, there are less than half as many cases where both the kinematic and combined models agree correctly, but the photometric model incorrectly disagrees ($37$ and $24$). These results show that while the completeness of the combined model is improved by incorporating the kinematic data alongside photometry, the relative potential for information gain by the combined model from the kinematic data is less than from the photometric data. The intersecting sector with the smallest population for both post-mergers and controls is that for which the photometry and kinematics correctly agree, but the combined model incorrectly disagrees ($15$ and $17$). This result demonstrates the rarity of cases for which the combined model fails to combine photometric and kinematic data which separately would have yielded a correct classification. 

Figure \ref{fig:phokin_agreement} shows examples of agreement and disagreement between the photometric and the kinematic models for post-merger galaxies. The decision thresholds for our models is always $P(\mathrm{PM})=0.5$, as justified by the calibration curves shown Figure \ref{fig:metrics}. The upper left quadrant of Figure \ref{fig:phokin_agreement} shows post-merger galaxies for which both models agree on their post-merger status -- many of which with a similar degree of confidence, $P(\mathrm{PM})$. Each case is visually reconcilable from either the photometry or velocity map (with the possible exception of the velocity map in the upper right panel of the quadrant). The lower right quadrant shows examples where the models incorrectly agree in favour of control classifications. Several such galaxies have regular velocity maps and morphologies. Others are more puzzling such as the velocity maps for the upper row of post-mergers in the quadrant. These misclassifications likely owe to the diverse and occasionally irregular morphological and kinematic structures represented in the control sample which do not originate from mergers. 

The lower left quadrant shows cases in which incorporating the kinematic information in addition to photometry is expected to be useful for improving post-merger selection completeness. These are cases in which the photometry incorrectly classifies post-mergers as controls but the kinematics make the correct classification. Indeed, many such cases are galaxies with low $R_1$ from Figure \ref{fig:mosaics} due to the high sensitivity of model metrics in the $R_1<40$ range. However, as noted in Section \ref{sec:sens_r1}, this higher completeness with the kinematic model is at the expense of reduced purity. The left panel of Figure \ref{fig:metrics} shows that combining the kinematics and photometry does not yield improved performance over the photometry in this regime. Therefore, the reduced purity in the kinematic model make its utility negligible as a complement to photometry in this $R_1$ regime. 

However, the lower left quadrant of Figure \ref{fig:phokin_agreement} also shows cases of post-merger galaxies whose morphologies are exceptionally normal, but whose kinematics are highly disturbed -- yielding negative classifications from photometry and positive classifications from kinematics. It may be expected (but not \emph{exactly}), for example, that post-merger galaxies whose average score between photometry and kinematics exceeds $\bar{P}_{pk}(\mathrm{PM})\geq0.5$ would be classified as a post-merger by the dual-input combined model. These are therefore, likely to be the post-merger galaxies contributing to the $1.6\%$ improvement in completeness of the combined model with respect to photometry alone -- while maintaining the same $93.5\%$ purity.


\bsp	
\label{lastpage}
\end{document}